\documentclass[aps,prl,reprint,superscriptaddress]{revtex4-1}

\usepackage{graphicx} 
\usepackage{color}

\begin{document}


\title{Validity of Kohler's rule in the pseudogap phase of the cuprate superconductors}


\author{M. K. Chan}
\email[]{mchan@physics.umn.edu}
\affiliation{School of Physics and Astronomy, University of Minnesota, Minneapolis, Minnesota 55455, USA}

\author{M. J. Veit}
\affiliation{School of Physics and Astronomy, University of Minnesota, Minneapolis, Minnesota 55455, USA}
\author{C. J. Dorow}
\altaffiliation{Present address: Department of Physics, University of California, San Diego, 9500 Gilman Drive La Jolla, CA 92093, USA}
\affiliation{School of Physics and Astronomy, University of Minnesota, Minneapolis, Minnesota 55455, USA}
\author{Y. Ge}
\affiliation{School of Physics and Astronomy, University of Minnesota, Minneapolis, Minnesota 55455, USA}
\author{Y. Li}
\affiliation{School of Physics and Astronomy, University of Minnesota, Minneapolis, Minnesota 55455, USA}
\author{W. Tabis}
\affiliation{School of Physics and Astronomy, University of Minnesota, Minneapolis, Minnesota 55455, USA}
\affiliation{AGH University of Science and Technology, Faculty of Physics and Applied Computer Science, 30-059 Krakow, Poland}
\author{Y. Tang}
\affiliation{School of Physics and Astronomy, University of Minnesota, Minneapolis, Minnesota 55455, USA}
\author{X. Zhao}
\affiliation{School of Physics and Astronomy, University of Minnesota, Minneapolis, Minnesota 55455, USA}
\affiliation{State Key Lab of Inorganic Synthesis and Preparative Chemistry, College of Chemistry, Jilin University, Changchun 130012, China}



\author{N. Bari\v{s}i\'c}
\email{nbarisic@ifs.hr}
\affiliation{School of Physics and Astronomy, University of Minnesota, Minneapolis, Minnesota 55455, USA}
\affiliation{Service de Physique de l'Etat Condens\'e, CEA-DSM-IRAMIS, F 91198 Gif-sur-Yvette, France}
\affiliation{Institute of Solid State Physics, Vienna University of Technology, 1040 Vienna, Austria}

\author{M. Greven}
\email{greven@physics.umn.edu}
\affiliation{School of Physics and Astronomy, University of Minnesota, Minneapolis, Minnesota 55455, USA}

\date{\today}
\begin{abstract}
We report in-plane resistivity ($\rho$) and transverse magnetoresistance (MR) measurements in underdoped HgBa$_2$CuO$_{4+\delta}$ (Hg1201). Contrary to the longstanding view that Kohler's rule is strongly violated in underdoped cuprates, we find that it is in fact satisfied in the pseudogap phase of Hg1201. The transverse MR shows a quadratic field dependence, $\delta\rho/\rho_o=a H^{2}$, with $a(T)\propto T^{-4}$. In combination with the observed $\rho\propto T^2$ dependence, this is consistent with a single Fermi-liquid quasiparticle scattering rate. We show that this behavior is universal, yet typically masked in cuprates with lower structural symmetry or strong disorder effects.
\end{abstract}


\maketitle

The unusual metallic `normal state' of the cuprates, from which superconductivity evolves upon cooling, has remained an enigma. A number of atypical observations seemingly at odds with the conventional Fermi-liquid theory of metals have been made particularly in the strange-metal regime above the pseudogap (PG) temperature $T^*$ (inset of Fig. 1(b))~\cite{hussey08}. In this regime, the in-plane resistivity exhibits an anomalous extended linear temperature dependence, $\rho \propto T$~\cite{martin90}, and the Hall effect is often described as $R_{\rm H} \propto 1/T$~\cite{hwang94,ando04}. In order to account for this anomolous behavior without abandoning a Fermi-liquid formalism, some descriptions have been formulated based on a  scattering rate whose magnitude varies around the in-plane Fermi surface, for example due  to anisotropic Umklapp scattering or coupling to a bosonic mode~\cite{hussey08} (e.g., spin~\cite{monthoux94} or charge~\cite{castellani95} fluctuations). Prominent non-Fermi-liquid prescriptions with far-ranging implications for the cuprate phase-diagram, such as the two-lifetime picture~\cite{anderson91} and the marginal-Fermi-liquid~\cite{varma89} have also been put forth. The former implies charge-spin separation while the latter is a signature of a proximate quantum critical point.

The transport behavior in the PG state ($T<T^*$) seems to be even less clear. Interpretation of this regime has been complicated not only because of the opening of the PG along portions of the Fermi surface, but also due to possible superconducting~\cite{xu00,*emery95}, antiferromagnetic~\cite{batlogg96,monthoux94}, and charge-spin stripe fluctuations~\cite{emery97}, which might influence transport properties. Electrical transport for temperatures below $T$* therefore has been generally described as a deviation from the better-behaved high-temperature behavior~\cite{hussey08}. 

Recent developments, however, suggests that $T^*$ marks a phase transition~\cite{shekhter13} into a state with broken time-reversal symmetry~\cite{fauque06,li08}. Additionally, the measurable extent of superconducting fluctuations is likely limited to only a rather small temperature range ($\approx 30$ K) above $T_{\rm c}$~\cite{grbic09,*bilbro11,*grbic11,*yu12,albenque07}. These strong indications that the PG regime is indeed a distinct phase calls for a clear description of its intrinsic properties. In fact, a simple $\rho =A_2 T^2$ dependence was recently reported for underdoped HgBa$_2$CuO$_{4+ \delta}$ (Hg1201)~\cite{barisic13}. It was also found that this Fermi-liquid-like behavior universally appears below a characteristic temperature $T^{**}$ ($T_{\rm c}<T^{**}<T^*$; see inset of Fig. 1(b))  and that the coefficient $A_2$, when normalized by the number of CuO$_2$ layers per unit cell, is also universal~\cite{barisic13}. This assertion of Fermi-liquid transport was further supported by  optical conductivity measurements that demonstrated an $\omega^2$ dependence of the scattering rate~\cite{mirzaei13}. 

\indent For a conventional metal, the change in isothermal resistivity $\delta\rho$ in an applied magnetic field ($H$) obeys a functional relation known as Kohler's rule: $\delta\rho/\rho_0=F(H/\rho_0)$, where $\rho_0$ is the zero-field resistivity at a given temperature. This relation follows from the fact that the magnetic field enters Boltzmann's equation in the combination $(H\tau)$ and that $\rho_0$ is inversely proportional to $\tau$, where $\tau$ is the scattering time. In the weak-field limit, most simple metals exhibit a $H^2$ dependence of the MR, so $\delta \rho/\rho_0 \propto \tau^2H^2$. A plot of $\delta\rho/\rho_0$ versus $(H/\rho_0)^2$ is expected to collapse to a single temperature-independent curve, as long as the number of carriers contributing to the transport is constant. Additionally, the electrical transport must also be described by a single scattering rate, or by several scattering rates whose relative contributions remain unchanged. For the case of a Fermi liquid with pure $1/\tau\propto T^2$, Kohler's rule is valid if $\delta \rho/\rho_0 \propto H^2T^{-4}$. 

For the cuprates, prior studies (e.g., in La$_{2-x}$Sr$_x$CuO$_4$ (LSCO) and YBa$_2$Cu$_3$O$_{6+y}$ (YBCO)) report that Kohler$'$s rule is strongly violated over a wide temperature range encompassing the PG phase~\cite{harris95,semba97,kimura96}. The implication of these results is that electrical transport in the cuprates is not as simple as implied by the recent zero-field dc and optical conductivity work~\cite{barisic13,mirzaei13}. 

In this Letter, we revisit the seemingly anomalous magneto-transport in the PG phase through in-plane resistivity and magnetoresistance measurements of Hg1201. Hg1201 has a simple tetragonal (P4/mmm) crystal structure with one copper-oxygen layer per primitive cell and features the highest $T_{\rm c}$ at optimal doping of all single-layer cuprates~\cite{eisaki04,putilin93}. Together with the availability of high-quality single crystals~~\cite{zhao06, barisic08, li11, barisic13b, leyraud13}, this makes Hg1201 a particularly interesting compound for transport studies. We demonstrate that Kohler's rule is in fact satisfied in the PG phase of Hg1201 and that the temperature dependences of $\rho$ and MR in the PG phase are consistent with a quasiparticle scattering rate expected from conventional Fermi-liquid theory. Importantly, we furthermore demonstrate that the same conclusion is valid for other members of the cuprate family if the intrinsic properties of the system are properly identified. 

\indent The preparation of Hg1201 samples for transport measurements was described previously in Refs.~\cite{zhao06,barisic08}. We present measurements on three samples: two with $T_{\rm c} = 70$ K (labeled HgUD70a and HgUD70b; hole doping $p\approx0.095$) and one with $T_{\rm c} = 81$ K (HgUD81; $p\approx0.11$), where the quoted hole concentrations are based on thermoelectric power measurements~\cite{yamamoto00}.
\begin{figure}[t!]
\includegraphics[width=.48\textwidth]{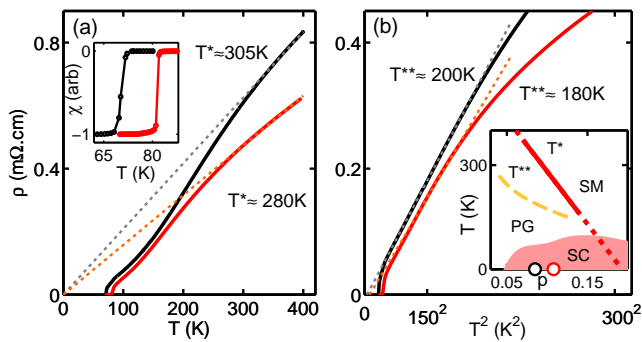}
\caption{ (a) Temperature dependence of the in-plane resistivity of two underdoped Hg1201 samples. Dotted lines are linear fits to the high-temperature behavior. Inset: Magnetic susceptibility shows the superconducting transition temperature $T_{\rm c} = 70\pm 1$K and $80.5\pm0.5$K for the two samples HgUD70a (black) and HgUD81 (red). The quoted $T_{\rm c}$ values are defined as the midpoint of the transition and the uncertainties correspond to 90$\%$ of the transition width. (b) Resistivity plotted versus $T^2$. Dotted lines are fits to the quadratic form $\rho=A_2 T^2$. Inset: Schematic temperature-hole doping phase diagram. The superconducting (SC), strange metal and pseudogap (PG) phases as well as the characteristic temperatures $T^*$ and $T^{**}$ are indicated. The circles represent the two doping levels for which data are presented in this Letter. \label{f1}}
\end{figure}
Figure 1(a) shows the temperature dependence of $\rho$. $T^*$ is determined from the deviation from approximate high-temperature linear behavior and agrees with prior reported values~\cite{li08, barisic13}. The same data are plotted versus $T^2$ in Fig. 1(b). Pure $\rho=A_2 T^2$ behavior is observed between the characteristic temperature $T^{**}$ and $\sim T_c + 20$ K. Both the linear and the quadratic dependence extrapolate to a negligible residual resistivity ($\rho_{\rm res} \approx 0$), which attests to the high quality of the crystals. 

\begin{figure*}
\includegraphics[width=\textwidth]{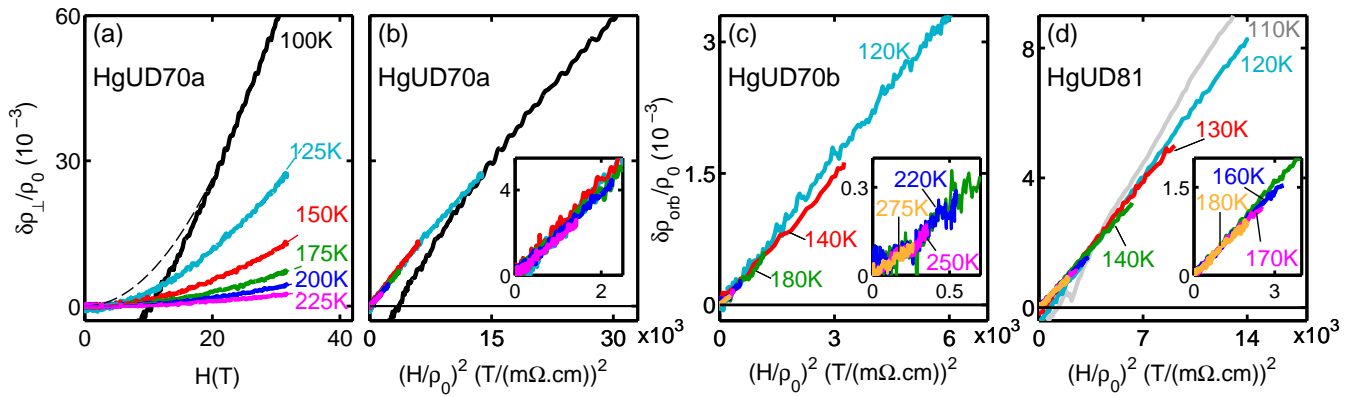}%
\caption{(a) Transverse MR with $H\parallel c$ and $j \parallel ab$ and (b) the corresponding Kohler plot for sample HgUD70a ($T_{\rm c} = 70$~K) measured in a maximum pulsed field of $30$~T. $\rho_0$ is defined as the zero-field resistivity extrapolated from fitting the data above 20~T to the form $\rho= \rho_0 +a^\prime H^2$. (c) Kohler plot for the orbital MR ($\rho(H\parallel c)- \rho (H \parallel j)$ with $j\parallel ab$) of  HgUD70b measured in a Quantum Design Inc. PPMS system up to 9 T. (d) Kohler plot of the orbital MR for HgUD81 measured in fields up to 9~T. Insets of (b) - (d) are low-field views of the respective panels. \label{kf2}}
\end{figure*}

\indent The magnetic field dependence of $\rho$ is measured in static and pulsed magnetic fields, as shown in Fig. \ref{kf2}~\cite{nhmfl}. The measurement in a pulsed field (up to $30$~T) was performed at LNCMI-Toulouse, France, in transverse geometry ($j{\parallel}ab, H{\parallel}c$). The dashed lines in Fig.~\ref{kf2}a are fits to $\delta\rho_\perp/\rho_0=a_{\perp}H^2$, where $a_\perp$ is the transverse MR coefficient. The large field range ensures high-quality fits. The Kohler plot of the data in Fig.~\ref{kf2}(a) is shown in Fig.~\ref{kf2}(b). Kohler$'$s rule is perfectly satisfied at all fields for temperatures between 125 K and 225 K, despite a change in $\rho_0$ by a factor of $\sim6$ in this temperature range. At 100 K, notwithstanding the deviation at low fields, the high-field data (where superconducting fluctuations~\cite{cooper09,albenque07,koka06} are suppressed, thus revealing the underlying normal state transport) exhibit the same slope. 

\indent In YBCO and LSCO, particularly at low doping, a longitudinal MR $\delta\rho_\parallel/\delta\rho_0$ ($j{\parallel}ab$, $H{\parallel}j$) is observed~~\cite{harris95, ando02}. This has been attributed to an isotropic spin-dependent term, which is excluded in the orbital MR defined as $\delta\rho_{\text{orb}}/\delta\rho_0\equiv (\delta\rho_\perp-\delta\rho_\parallel)/\rho_0$~\cite{harris95}. To test the possibility that such contributions might affect our result, we measured the second $T_c = 70$~K sample (HgUD70b) and established that longitudinal MR is at least an order of magnitude smaller. A similiarly small longitudinal MR is observed in other cuprates close to optimal-doping~\cite{kimura96,ando02}. To demonstrate the robustness of Kohler's rule, the orbital MR for HgUD70b is shown in Fig. 2(c). We find that $\delta\rho_{\text{orb}}/\delta\rho_0$ also satisfies Kohler$'$s rule from $125$~K to $275$~K. Since the longitudinal contribution is small, the transverse and orbital MR coefficients of HgUD70a and HgUD70b, respectively, are indistinguishable, as shown in Fig. 3(a). Kohler's rule is also found to be obeyed in the PG phase of HgUD81 (Fig.2(d)). Our result is therefore not isolated to a particular doping level. 

\begin{figure}[b]
\includegraphics[width=.48\textwidth]{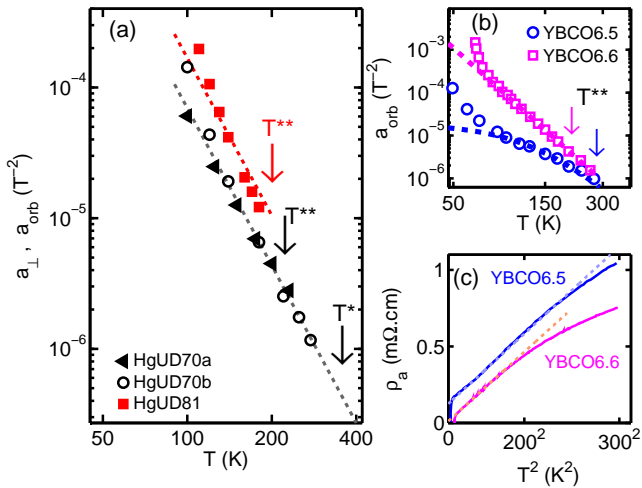}%
\caption{(a) Temperature dependence of the orbital MR coefficient $a_{\text{orb}} \equiv(a_\perp-a_\parallel)$ for HgUD70b and HgUD81 and of the transverse MR coefficient $a_\perp$ for HgUD70a. The dashed lines are fits to $a=(b T)^{-4}$. The arrows mark the characteristic temperatures $T^*$ and $T^{**}$. (b) and (c) show representative $a_{\text{orb}}$ and $a$-axis in-plane resistivity $\rho_a$ from Ref.~\cite{ando02} for YBCO6.5 (YBa$_2$Cu$_3$O$_{6+y}$ with $y=0.5$, $T_c\approx35$~K, $p\approx 0.073$) and YBCO6.6 ($y=0.6$, $T_c\approx50$~K, $p\approx 0.085$). We estimate the doping levels by comparing the quoted $T_{\rm c}$ values to Ref.~~\cite{liang06}. Dashed lines in (b) are fits to the form $(c+bT^2)^{-2}$, with $c=0$ for YBCO6.6. The resistivity in (c) is plotted versus $T^2$ to highlight the $\rho_a\propto T^2$ behavior indicated by dashed lines below the characteristic temperature $T^{**}$, consistent with the inset of Fig. 1(b). \label{fa}}
\end{figure}


\indent As shown in Fig. 3(a), the MR coefficient ($a_{\perp\rm{,orb}}=\delta\rho_{\perp\rm{,orb}}/\rho_o  H^2$) exhibits $T^{-4}$ dependence that extends from approximately $100$~K to at least $T^{**} \approx 180$~K in HgUD81 and even beyond $T^{**} \approx 200$~K in HgUD70b. Since $\delta\rho/\rho_0\propto H^2\tau^2\propto H^2 T^{-4}$, it follows that $1/\tau\propto T^2$. For Hg1201, this is consistent with the Fermi-liquid scattering rate below $T^{**}$ infered from the temperature dependence of $\rho$.


\indent The $T^{-4}$ dependence of the MR has previously been reported for a number of cuprates~\cite{harris95, ando04,kimura96}. Nevertheless, Kohler$'$s rule was claimed to be violated throughout most of the phase diagram~\cite{harris95,semba97,ando04,kimura96}. Prior conclusions pertaining to the violation of Kohler's rule in the PG phase of YBCO can be attributed to the difficulty of measuring the underlying pure $\rho\propto T^2$ behavior. In YBCO, Cu-O chains form along the crystallographic $b$ direction, which contribute to the electrical transport and could prevent a clean measurement of the intrinsic resistivity of the  CuO$_2$ planes. Since the relative contribution to $\rho$ from the chains is temperature dependent, the combined contributions would violate Kohler's rule and could be a contributing factor to the negative result for twinned crystals reported in Ref.~\cite{harris95}. Measurements of very underdoped non-superconducting tetragonal YBCO ($p\approx0.03$)~\cite{ando04}, which does not feature CuO chains, and of the $a$-axis resistivity $\rho_a$ in detwinned YBCO crystals at higher doping~\cite{ando02,lee05} (the chains are not expected to contribute to transport perpendicular to them) have, in fact, revealed a $T^2$ resistivity. For example, Fig. \ref{fa}(c) shows representative data from Ref.~\cite{ando02} with $T^2$-dependence below a characteristic temperature that decreases with increasing doping, consistent with $T^{**}$ (Fig. 1(b) inset)~\cite{barisic13}. 

As shown in Fig. \ref{fa}(b), YBCO6.6 exhibits the expected $a_{orb}\propto T^{-4}$ dependence of the MR.  In conjunction with the $\rho=A_2 T^2$ dependence, we conclude that YBCO6.6 obeys Kohler's rule below $T^{**}$. In slightly more underdoped YBCO6.5, on the other hand, $\rho=\rho_{\text res}+A_2 T^2$ with a large  residual resistivity $\rho_{\text res}$. This is reflected in the MR, which is fit to $a_{\text{orb}} = (c+bT^2)^{-2}$~\cite{harris95,ando02} (Fig. 3(c)), where $c$ is a residual temperature independent contribution to the scattering rate~\cite{harris95}. We note that the ratio between the residual term and the $T^2$ coefficient manifested in the MR ($c/b=9700~\text{K}^2$) and in the zero-field resistivity ($\rho_{\text res}/A_2=10800~\text{K}^2$) are equivalent to about 10\%, which is approximately the uncertainty of the fitted coefficients, thus supporting our conclusion that the same residual scattering rate enters both quantities. Consequently, Kohler's rule is violated in YBCO6.5 solely due to the contribution of a residual temperature independent scattering rate. Our main conclusion is that, notwithstanding the significant differences in crystal structures, the normal state below $T^{**}$ in the cuprates is universally characterized by single, isotropic scattering rate consistent with Fermi-liquid transport. 

\begin{figure}
\includegraphics[width=.48\textwidth]{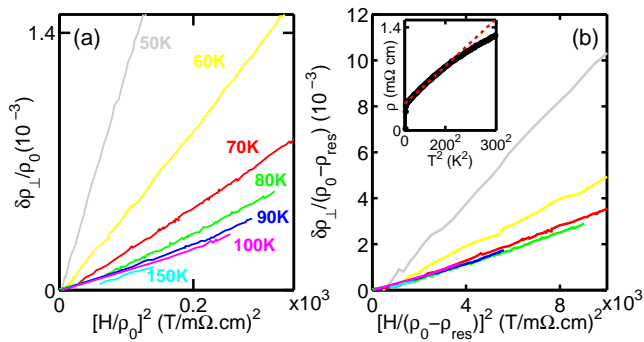}%
\caption{(a) Kohler's plot for  LSCO $x=0.09$ from Ref.~\cite{kimura96} (b) Modified Kohler's plot, with  $\rho_0$ replaced by $\rho_0-\rho_{\text{res}}$. Inset: Temperature dependence of the planar resistivity $\rho$ in zero field. The dotted red line is a fit to $\rho=\rho_{\text{res}}+A_2T^2$. $\rho_{\text{res}}$ is determined from extrapoating to $T=0$. \label{flsco}}
\end{figure}

\indent LSCO exhibits some transport properties that are at variance with YBCO and Hg1201: a particularly large residual resistivity and a tendency toward insulating low-temperature behavior~\cite{boebinger96} below optimal doping, instead of the metallic behavior and quantum oscillations found in YBCO and Hg1201~\cite{leyraud07, leboeuf11, barisic13b, leyraud13}. Both LSCO and Hg1201 are single-layer cuprates. Yet LSCO features lower structural symmetry, more disorder and a significantly lower optimal $T_{\rm c}$ ($\approx 40$~K) than simple tetragonal Hg1201~\cite{eisaki04}. Nevertheless, the planar resistivity for moderately underdoped LSCO $x=0.09$ is fit well to $\rho=\rho_{\text{res}}+A_2 T^2$ between $70$~K$-200$~K with a large $\rho_{\text{res}}$ (inset of Fig. \ref{flsco}(b))~\cite{kimura96}. This is consistent with  $T^{**}\approx 200$~K indicated in Ref.~\cite{barisic13}. Fig.~\ref{flsco}(a) shows the strong violation of Kohler's rule in this sample. However, in Fig.~\ref{flsco}(b), we show that the MR curves between $70$~K and $150$~K collapse after subtracting $\rho_{\text{res}}$ from $\rho_0$. This is a suprising result, because it implies that the residual scattering rate does not affect the MR $\delta\rho$. This property is unlike that of YBCO6.5 highlighted above, where the residual scattering rate enters both $\rho$ and $\delta\rho/\rho_0$ equally. The deviation from Kohler's rule below 70~K might be related to the observation of a large Nernst effect signal and could be attributed to superconducting fluctuations~\cite{wang06} or incipient stripe order~\cite{cyr09}. 

The anomalous behavior of $\rho_{\text{res}}$ in LSCO suggestes that it is not solely due to electron-impurity scattering. Indeed, upon decreasing the hole concentration in LSCO, either chemically~\cite{ando04b} or through electrostatic gating~\cite{bollinger11} (which does not change the impurity concentration), $\rho_{\text res}$ extrapolated from high temperatures increases progressively upon approaching the superconductor-insulator phase transition. Electrostatic gating revealed a critical CuO$_2$ sheet resistance of  $R_c={\rm h}/(2{\rm e})^2=6.5 \rm{k\Omega}$~\cite{bollinger11}. Similiar observations for the superconductor-insulator transition have been made for YBCO~\cite{walker95,leng11} and Bi$_2$Sr$_2$Y$_x$Ca$_{1-x}$Cu$_2$O$_8$~\cite{mandrus91}. Furthermore, for LSCO, a non-zero extrapolated $\rho_{\text res}$ is observed in superconducting samples up to optimal doping~\cite{ando04b}, and when the superconductivity is surpressed in a sufficiently large magnetic field, an insulating ground state is revealed~\cite{boebinger96}. This phenomenon coincides with the presence of nearly static incommensurate spin correlations observed with neutron scattering~\cite{yamada98,kofu09}. In contrast, $\rho_{\text{res}}$ for Hg1201 is negligible even in the most underdoped single crystals measured ($p\approx 0.055$, $T_c=45$~K~\cite{barisic13}). Finally, despite the compound-dependent variation of $\rho_{\text{res}}$, the high-temperature $\rho=A_2T^2$ dependence below $T^{**}$  established in Ref.~\cite{barisic13} is universal. 

\indent The emerging picture for electrical transport in the underdoped cuprates at temperatures below $T^{**}$ is that of a Fermi liquid, characterized by a $T^2$ and $\omega^2$~\cite{mirzaei13} quasiparticle scattering rate. Angle-resolved photoemission spectroscopy (ARPES) indicates that a large PG appears in the antinodal density of states below $T^*$, leaving small arcs around the nodal points~\cite{damascelli04}. The Fermi liquid must therefore reside on the arcs~\cite{ando04,lee05,barisic13,gorkov13}, where quasiparticle peaks have been detected with ARPES~\cite{yoshida03}. Upon warming above $T^{**}$, the resistivity deviates from the simple quadratic temperature dependence, yet we observe that the MR continues to follow $\sim T^{-4}$ behavior, unperturbed through $T^{**}$. Whether or not $T^{**}$ is a true phase transition or merely a crossover temperature, possibly marking the temperature below which the pseudogap is fully formed, is still an open question. 

One consequence of the validity of Kohler's rule demonstrated in the present work is that the Fermi surface should remain largely temperature independent between $T^{**}$ and $T_{\rm c}$. Charge-density-wave (CDW) correlations have recently been observed in underdoped YBCO~\cite{ghiringhelli12,chang12}, and also recently in Hg1201 at the same hole concentration as the HgUD70 samples studied here~\cite{tabis14}. Interestingly, the onset of these correlations in both Hg1201 and YBCO appear to correlate with $T^{**}$~\cite{tabis14}. However, the Fermi-liquid regime below $T^{**}$ extends to very low hole concentrations~\cite{barisic13}, in contrast to the CDW order, which appears to be tied to the doping range of the $T_c( p )$ plateau~\cite{ghiringhelli12}. Moreover, resistivity~\cite{barisic13} and ARPES~\cite{zhao03} results suggest the existence of an arc-like surface with a doping independent Fermi velocity. The appearence of CDW correlations might therefore be simply contingent on the stable Fermi surface below $T^{**}$ suggested in this work. This is consistent with the notion that the CDW wave-vector connects parts of the Fermi arcs in the PG state and is temperature independent~\cite{comin14,dasilva14,barisic09}. ARPES results do indicate that the arc length remains constant over a wide temperature range in the PG regime of (Bi,Pb)$_2$(Sr,La)$_2$CuO$_{6+\delta}$ (Bi2201) and Bi$_2$Sr$_2$CaCu$_2$O$_{8+\delta}$ (Bi2212) near optimal doping~\cite{kondo13}. It would be interesting to extend such measurements to lower hole concentrations concentrations, and to Hg1201, for which quantitative ARPES measurements have recently become feasible~\cite{vishik14}.

\indent We thank C. Proust and B. Vignolle for technical assistance in performing the pulsed field measurements at LNCMI-Toulouse, France. We also acknowledge technical assistance by Jan Jaroszynski at the NHMFL, Tallahassee, FL, USA, while performing high static field measurements. The work at the University of Minnesota was supported by the Department of Energy, Office of Basic Energy Sciences, under Award No. DE-SC0006858. The work in Toulouse was supported by the French ANR SUPERFIELD, Euromagnet II, and the LABEX NEXT. N.B. acknowledges support though a Marie Curie Fellowship and European Research Council (Advanced Grant Quantum Puzzle, no. 227378). 

\bibliography{Kohlers_MC15}

\begin{thebibliography}{63}%
\makeatletter
\providecommand \@ifxundefined [1]{%
 \@ifx{#1\undefined}
}%
\providecommand \@ifnum [1]{%
 \ifnum #1\expandafter \@firstoftwo
 \else \expandafter \@secondoftwo
 \fi
}%
\providecommand \@ifx [1]{%
 \ifx #1\expandafter \@firstoftwo
 \else \expandafter \@secondoftwo
 \fi
}%
\providecommand \natexlab [1]{#1}%
\providecommand \enquote  [1]{``#1''}%
\providecommand \bibnamefont  [1]{#1}%
\providecommand \bibfnamefont [1]{#1}%
\providecommand \citenamefont [1]{#1}%
\providecommand \href@noop [0]{\@secondoftwo}%
\providecommand \href [0]{\begingroup \@sanitize@url \@href}%
\providecommand \@href[1]{\@@startlink{#1}\@@href}%
\providecommand \@@href[1]{\endgroup#1\@@endlink}%
\providecommand \@sanitize@url [0]{\catcode `\\12\catcode `\$12\catcode
  `\&12\catcode `\#12\catcode `\^12\catcode `\_12\catcode `\%12\relax}%
\providecommand \@@startlink[1]{}%
\providecommand \@@endlink[0]{}%
\providecommand \url  [0]{\begingroup\@sanitize@url \@url }%
\providecommand \@url [1]{\endgroup\@href {#1}{\urlprefix }}%
\providecommand \urlprefix  [0]{URL }%
\providecommand \Eprint [0]{\href }%
\providecommand \doibase [0]{http://dx.doi.org/}%
\providecommand \selectlanguage [0]{\@gobble}%
\providecommand \bibinfo  [0]{\@secondoftwo}%
\providecommand \bibfield  [0]{\@secondoftwo}%
\providecommand \translation [1]{[#1]}%
\providecommand \BibitemOpen [0]{}%
\providecommand \bibitemStop [0]{}%
\providecommand \bibitemNoStop [0]{.\EOS\space}%
\providecommand \EOS [0]{\spacefactor3000\relax}%
\providecommand \BibitemShut  [1]{\csname bibitem#1\endcsname}%
\let\auto@bib@innerbib\@empty
\bibitem [{\citenamefont {Hussey}(2008)}]{hussey08}%
  \BibitemOpen
  \bibfield  {author} {\bibinfo {author} {\bibfnamefont {N.~E.}\ \bibnamefont
  {Hussey}},\ }\href@noop {} {\bibfield  {journal} {\bibinfo  {journal} {J.
  Phys. Condens. Matter}\ }\textbf {\bibinfo {volume} {20}},\ \bibinfo {pages}
  {123201} (\bibinfo {year} {2008})}\BibitemShut {NoStop}%
\bibitem [{\citenamefont {Martin}\ \emph {et~al.}(1990)\citenamefont {Martin},
  \citenamefont {Fiory}, \citenamefont {Fleming}, \citenamefont {Schneemeyer},\
  and\ \citenamefont {Waszczak}}]{martin90}%
  \BibitemOpen
  \bibfield  {author} {\bibinfo {author} {\bibfnamefont {S.}~\bibnamefont
  {Martin}}, \bibinfo {author} {\bibfnamefont {A.~T.}\ \bibnamefont {Fiory}},
  \bibinfo {author} {\bibfnamefont {R.~M.}\ \bibnamefont {Fleming}}, \bibinfo
  {author} {\bibfnamefont {L.~F.}\ \bibnamefont {Schneemeyer}}, \ and\ \bibinfo
  {author} {\bibfnamefont {J.~V.}\ \bibnamefont {Waszczak}},\ }\href@noop {}
  {\bibfield  {journal} {\bibinfo  {journal} {Phys. Rev. B}\ }\textbf {\bibinfo
  {volume} {41}},\ \bibinfo {pages} {846} (\bibinfo {year} {1990})}\BibitemShut
  {NoStop}%
\bibitem [{\citenamefont {Hwang}\ \emph {et~al.}(1994)\citenamefont {Hwang},
  \citenamefont {Batlogg}, \citenamefont {Takagi}, \citenamefont {Kao},
  \citenamefont {Kwo}, \citenamefont {Cava}, \citenamefont {Krajewski},\ and\
  \citenamefont {W.~F.~Peck}}]{hwang94}%
  \BibitemOpen
  \bibfield  {author} {\bibinfo {author} {\bibfnamefont {H.}~\bibnamefont
  {Hwang}}, \bibinfo {author} {\bibfnamefont {B.}~\bibnamefont {Batlogg}},
  \bibinfo {author} {\bibfnamefont {H.}~\bibnamefont {Takagi}}, \bibinfo
  {author} {\bibfnamefont {H.~L.}\ \bibnamefont {Kao}}, \bibinfo {author}
  {\bibfnamefont {J.}~\bibnamefont {Kwo}}, \bibinfo {author} {\bibfnamefont
  {R.~J.}\ \bibnamefont {Cava}}, \bibinfo {author} {\bibfnamefont {J.~J.}\
  \bibnamefont {Krajewski}}, \ and\ \bibinfo {author} {\bibfnamefont
  {J.}~\bibnamefont {W.~F.~Peck}},\ }\href@noop {} {\bibfield  {journal}
  {\bibinfo  {journal} {Phys. Rev. Lett.}\ }\textbf {\bibinfo {volume} {72}},\
  \bibinfo {pages} {2636} (\bibinfo {year} {1994})}\BibitemShut {NoStop}%
\bibitem [{\citenamefont {Ando}\ \emph
  {et~al.}(2004{\natexlab{a}})\citenamefont {Ando}, \citenamefont {Kurita},
  \citenamefont {Komiya}, \citenamefont {Ono},\ and\ \citenamefont
  {Segawa}}]{ando04}%
  \BibitemOpen
  \bibfield  {author} {\bibinfo {author} {\bibfnamefont {Y.}~\bibnamefont
  {Ando}}, \bibinfo {author} {\bibfnamefont {Y.}~\bibnamefont {Kurita}},
  \bibinfo {author} {\bibfnamefont {S.}~\bibnamefont {Komiya}}, \bibinfo
  {author} {\bibfnamefont {S.}~\bibnamefont {Ono}}, \ and\ \bibinfo {author}
  {\bibfnamefont {K.}~\bibnamefont {Segawa}},\ }\href@noop {} {\bibfield
  {journal} {\bibinfo  {journal} {Phys. Rev. Lett.}\ }\textbf {\bibinfo
  {volume} {92}},\ \bibinfo {pages} {197001} (\bibinfo {year}
  {2004}{\natexlab{a}})}\BibitemShut {NoStop}%
\bibitem [{\citenamefont {Monthoux}\ and\ \citenamefont
  {Pines}(1994)}]{monthoux94}%
  \BibitemOpen
  \bibfield  {author} {\bibinfo {author} {\bibfnamefont {P.}~\bibnamefont
  {Monthoux}}\ and\ \bibinfo {author} {\bibfnamefont {D.}~\bibnamefont
  {Pines}},\ }\href@noop {} {\bibfield  {journal} {\bibinfo  {journal} {Phys.
  Rev. B}\ }\textbf {\bibinfo {volume} {49}},\ \bibinfo {pages} {4261}
  (\bibinfo {year} {1994})}\BibitemShut {NoStop}%
\bibitem [{\citenamefont {Castellani}\ \emph {et~al.}(1995)\citenamefont
  {Castellani}, \citenamefont {Di~Castro},\ and\ \citenamefont
  {Grilli}}]{castellani95}%
  \BibitemOpen
  \bibfield  {author} {\bibinfo {author} {\bibfnamefont {C.}~\bibnamefont
  {Castellani}}, \bibinfo {author} {\bibfnamefont {C.}~\bibnamefont
  {Di~Castro}}, \ and\ \bibinfo {author} {\bibfnamefont {M.}~\bibnamefont
  {Grilli}},\ }\href@noop {} {\bibfield  {journal} {\bibinfo  {journal} {Phys.
  Rev. Lett.}\ }\textbf {\bibinfo {volume} {75}},\ \bibinfo {pages} {4650}
  (\bibinfo {year} {1995})}\BibitemShut {NoStop}%
\bibitem [{\citenamefont {Anderson}(1991)}]{anderson91}%
  \BibitemOpen
  \bibfield  {author} {\bibinfo {author} {\bibfnamefont {P.~W.}\ \bibnamefont
  {Anderson}},\ }\href@noop {} {\bibfield  {journal} {\bibinfo  {journal}
  {Phys. Rev. Lett.}\ }\textbf {\bibinfo {volume} {67}},\ \bibinfo {pages}
  {2092} (\bibinfo {year} {1991})}\BibitemShut {NoStop}%
\bibitem [{\citenamefont {Varma}(1989)}]{varma89}%
  \BibitemOpen
  \bibfield  {author} {\bibinfo {author} {\bibfnamefont {C.}~\bibnamefont
  {Varma}},\ }\href@noop {} {\bibfield  {journal} {\bibinfo  {journal} {Phys.
  Rev. Lett.}\ }\textbf {\bibinfo {volume} {63}},\ \bibinfo {pages} {1996}
  (\bibinfo {year} {1989})}\BibitemShut {NoStop}%
\bibitem [{\citenamefont {Xu}\ \emph {et~al.}(2000)\citenamefont {Xu},
  \citenamefont {Ong}, \citenamefont {Wang}, \citenamefont {Kakeshita},\ and\
  \citenamefont {Uchida}}]{xu00}%
  \BibitemOpen
  \bibfield  {author} {\bibinfo {author} {\bibfnamefont {Z.~A.}\ \bibnamefont
  {Xu}}, \bibinfo {author} {\bibfnamefont {N.~P.}\ \bibnamefont {Ong}},
  \bibinfo {author} {\bibfnamefont {Y.}~\bibnamefont {Wang}}, \bibinfo {author}
  {\bibfnamefont {T.}~\bibnamefont {Kakeshita}}, \ and\ \bibinfo {author}
  {\bibfnamefont {S.}~\bibnamefont {Uchida}},\ }\href@noop {} {\bibfield
  {journal} {\bibinfo  {journal} {Nature (London)}\ }\textbf {\bibinfo {volume}
  {406}},\ \bibinfo {pages} {486} (\bibinfo {year} {2000})}\BibitemShut
  {NoStop}%
\bibitem [{\citenamefont {Emery}\ and\ \citenamefont
  {Kivelson}(1995)}]{emery95}%
  \BibitemOpen
  \bibfield  {author} {\bibinfo {author} {\bibfnamefont {V.~J.}\ \bibnamefont
  {Emery}}\ and\ \bibinfo {author} {\bibfnamefont {S.~A.}\ \bibnamefont
  {Kivelson}},\ }\href@noop {} {\bibfield  {journal} {\bibinfo  {journal}
  {Nature (London)}\ }\textbf {\bibinfo {volume} {374}},\ \bibinfo {pages}
  {434} (\bibinfo {year} {1995})}\BibitemShut {NoStop}%
\bibitem [{\citenamefont {Batlogg}\ and\ \citenamefont
  {Emery}(1996)}]{batlogg96}%
  \BibitemOpen
  \bibfield  {author} {\bibinfo {author} {\bibfnamefont {B.}~\bibnamefont
  {Batlogg}}\ and\ \bibinfo {author} {\bibfnamefont {V.~J.}\ \bibnamefont
  {Emery}},\ }\href@noop {} {\bibfield  {journal} {\bibinfo  {journal} {Nature
  (London)}\ }\textbf {\bibinfo {volume} {382}},\ \bibinfo {pages} {20}
  (\bibinfo {year} {1996})}\BibitemShut {NoStop}%
\bibitem [{\citenamefont {Emery}\ \emph {et~al.}(1997)\citenamefont {Emery},
  \citenamefont {Kivelson},\ and\ \citenamefont {Zachar}}]{emery97}%
  \BibitemOpen
  \bibfield  {author} {\bibinfo {author} {\bibfnamefont {V.~J.}\ \bibnamefont
  {Emery}}, \bibinfo {author} {\bibfnamefont {S.~A.}\ \bibnamefont {Kivelson}},
  \ and\ \bibinfo {author} {\bibfnamefont {O.}~\bibnamefont {Zachar}},\
  }\href@noop {} {\bibfield  {journal} {\bibinfo  {journal} {Phys. Rev. B}\
  }\textbf {\bibinfo {volume} {56}},\ \bibinfo {pages} {6120} (\bibinfo {year}
  {1997})}\BibitemShut {NoStop}%
\bibitem [{\citenamefont {Shekhter}\ \emph {et~al.}(2013)\citenamefont
  {Shekhter}, \citenamefont {Ramshaw}, \citenamefont {Liang}, \citenamefont
  {Hardy}, \citenamefont {Bonn}, \citenamefont {Balakirev}, \citenamefont
  {McDonald}, \citenamefont {Betts}, \citenamefont {Riggs},\ and\ \citenamefont
  {Migliori}}]{shekhter13}%
  \BibitemOpen
  \bibfield  {author} {\bibinfo {author} {\bibfnamefont {A.}~\bibnamefont
  {Shekhter}}, \bibinfo {author} {\bibfnamefont {B.~J.}\ \bibnamefont
  {Ramshaw}}, \bibinfo {author} {\bibfnamefont {R.}~\bibnamefont {Liang}},
  \bibinfo {author} {\bibfnamefont {W.~N.}\ \bibnamefont {Hardy}}, \bibinfo
  {author} {\bibfnamefont {D.~A.}\ \bibnamefont {Bonn}}, \bibinfo {author}
  {\bibfnamefont {F.~F.}\ \bibnamefont {Balakirev}}, \bibinfo {author}
  {\bibfnamefont {R.~D.}\ \bibnamefont {McDonald}}, \bibinfo {author}
  {\bibfnamefont {J.~B.}\ \bibnamefont {Betts}}, \bibinfo {author}
  {\bibfnamefont {S.~C.}\ \bibnamefont {Riggs}}, \ and\ \bibinfo {author}
  {\bibfnamefont {A.}~\bibnamefont {Migliori}},\ }\href@noop {} {\bibfield
  {journal} {\bibinfo  {journal} {Nature (London)}\ }\textbf {\bibinfo {volume}
  {498}},\ \bibinfo {pages} {75} (\bibinfo {year} {2013})}\BibitemShut
  {NoStop}%
\bibitem [{\citenamefont {Fauqu\'{e}}\ \emph {et~al.}(2006)\citenamefont
  {Fauqu\'{e}}, \citenamefont {Sidis}, \citenamefont {Hinkov}, \citenamefont
  {Pailh\'{e}s}, \citenamefont {Lin}, \citenamefont {Chaud},\ and\
  \citenamefont {Bourges}}]{fauque06}%
  \BibitemOpen
  \bibfield  {author} {\bibinfo {author} {\bibfnamefont {B.}~\bibnamefont
  {Fauqu\'{e}}}, \bibinfo {author} {\bibfnamefont {Y.}~\bibnamefont {Sidis}},
  \bibinfo {author} {\bibfnamefont {V.}~\bibnamefont {Hinkov}}, \bibinfo
  {author} {\bibfnamefont {S.}~\bibnamefont {Pailh\'{e}s}}, \bibinfo {author}
  {\bibfnamefont {C.~T.}\ \bibnamefont {Lin}}, \bibinfo {author} {\bibfnamefont
  {X.}~\bibnamefont {Chaud}}, \ and\ \bibinfo {author} {\bibfnamefont
  {P.}~\bibnamefont {Bourges}},\ }\href@noop {} {\bibfield  {journal} {\bibinfo
   {journal} {Phys. Rev. Lett.}\ }\textbf {\bibinfo {volume} {96}},\ \bibinfo
  {pages} {197001} (\bibinfo {year} {2006})}\BibitemShut {NoStop}%
\bibitem [{\citenamefont {Li}\ \emph {et~al.}(2008)\citenamefont {Li},
  \citenamefont {Bal\`{e}dent}, \citenamefont {Bari\v{s}i\'c}, \citenamefont
  {Cho}, \citenamefont {Fauqu\`{e}}, \citenamefont {Sidis}, \citenamefont {Yu},
  \citenamefont {Zhao}, \citenamefont {Bourges},\ and\ \citenamefont
  {Greven}}]{li08}%
  \BibitemOpen
  \bibfield  {author} {\bibinfo {author} {\bibfnamefont {Y.}~\bibnamefont
  {Li}}, \bibinfo {author} {\bibfnamefont {V.}~\bibnamefont {Bal\`{e}dent}},
  \bibinfo {author} {\bibfnamefont {N.}~\bibnamefont {Bari\v{s}i\'c}}, \bibinfo
  {author} {\bibfnamefont {Y.}~\bibnamefont {Cho}}, \bibinfo {author}
  {\bibfnamefont {B.}~\bibnamefont {Fauqu\`{e}}}, \bibinfo {author}
  {\bibfnamefont {Y.}~\bibnamefont {Sidis}}, \bibinfo {author} {\bibfnamefont
  {G.}~\bibnamefont {Yu}}, \bibinfo {author} {\bibfnamefont {X.}~\bibnamefont
  {Zhao}}, \bibinfo {author} {\bibfnamefont {P.}~\bibnamefont {Bourges}}, \
  and\ \bibinfo {author} {\bibfnamefont {M.}~\bibnamefont {Greven}},\
  }\href@noop {} {\bibfield  {journal} {\bibinfo  {journal} {Nature}\ }\textbf
  {\bibinfo {volume} {455}},\ \bibinfo {pages} {372} (\bibinfo {year}
  {2008})}\BibitemShut {NoStop}%
\bibitem [{\citenamefont {Grbi\'{c}}\ \emph {et~al.}(2009)\citenamefont
  {Grbi\'{c}}, \citenamefont {Bari\v{s}i\'{c}}, \citenamefont {Dul\v{c}i\'{c}},
  \citenamefont {Kup\v{c}i\'{c}}, \citenamefont {Li}, \citenamefont {Zhao},
  \citenamefont {Yu}, \citenamefont {Dressel}, \citenamefont {Greven},\ and\
  \citenamefont {Po\v{z}ek}}]{grbic09}%
  \BibitemOpen
  \bibfield  {author} {\bibinfo {author} {\bibfnamefont {M.~S.}\ \bibnamefont
  {Grbi\'{c}}}, \bibinfo {author} {\bibnamefont {Bari\v{s}i\'{c}}}, \bibinfo
  {author} {\bibfnamefont {A.}~\bibnamefont {Dul\v{c}i\'{c}}}, \bibinfo
  {author} {\bibfnamefont {I.}~\bibnamefont {Kup\v{c}i\'{c}}}, \bibinfo
  {author} {\bibfnamefont {Y.}~\bibnamefont {Li}}, \bibinfo {author}
  {\bibfnamefont {X.}~\bibnamefont {Zhao}}, \bibinfo {author} {\bibfnamefont
  {G.}~\bibnamefont {Yu}}, \bibinfo {author} {\bibnamefont {Dressel}}, \bibinfo
  {author} {\bibfnamefont {M.}~\bibnamefont {Greven}}, \ and\ \bibinfo {author}
  {\bibfnamefont {M.}~\bibnamefont {Po\v{z}ek}},\ }\href@noop {} {\bibfield
  {journal} {\bibinfo  {journal} {Phys. Rev. B}\ }\textbf {\bibinfo {volume}
  {80}},\ \bibinfo {pages} {094511} (\bibinfo {year} {2009})}\BibitemShut
  {NoStop}%
\bibitem [{\citenamefont {Bilbro}\ \emph {et~al.}(2011)\citenamefont {Bilbro},
  \citenamefont {Aguilar}, \citenamefont {Logvenov}, \citenamefont {Pelleg},
  \citenamefont {Bo\v{z}ovi\'c},\ and\ \citenamefont {Armitage}}]{bilbro11}%
  \BibitemOpen
  \bibfield  {author} {\bibinfo {author} {\bibfnamefont {L.~S.}\ \bibnamefont
  {Bilbro}}, \bibinfo {author} {\bibfnamefont {R.~V.}\ \bibnamefont {Aguilar}},
  \bibinfo {author} {\bibfnamefont {G.}~\bibnamefont {Logvenov}}, \bibinfo
  {author} {\bibfnamefont {O.}~\bibnamefont {Pelleg}}, \bibinfo {author}
  {\bibfnamefont {I.}~\bibnamefont {Bo\v{z}ovi\'c}}, \ and\ \bibinfo {author}
  {\bibfnamefont {N.~P.}\ \bibnamefont {Armitage}},\ }\href@noop {} {\bibfield
  {journal} {\bibinfo  {journal} {Nat. Phys.}\ }\textbf {\bibinfo {volume}
  {7}},\ \bibinfo {pages} {298} (\bibinfo {year} {2011})}\BibitemShut {NoStop}%
\bibitem [{\citenamefont {Grbi\ifmmode~\acute{c}\else \'{c}\fi{}}\ \emph
  {et~al.}(2011)\citenamefont {Grbi\ifmmode~\acute{c}\else \'{c}\fi{}},
  \citenamefont {Po\ifmmode~\check{z}\else \v{z}\fi{}ek}, \citenamefont {Paar},
  \citenamefont {Hinkov}, \citenamefont {Raichle}, \citenamefont {Haug},
  \citenamefont {Keimer}, \citenamefont {Bari\ifmmode \check{s}\else
  \v{s}\fi{}i\ifmmode~\acute{c}\else \'{c}\fi{}},\ and\ \citenamefont
  {Dul\ifmmode \check{c}\else \v{c}\fi{}i\ifmmode~\acute{c}\else
  \'{c}\fi{}}}]{grbic11}%
  \BibitemOpen
  \bibfield  {author} {\bibinfo {author} {\bibfnamefont {M.~S.}\ \bibnamefont
  {Grbi\ifmmode~\acute{c}\else \'{c}\fi{}}}, \bibinfo {author} {\bibfnamefont
  {M.}~\bibnamefont {Po\ifmmode~\check{z}\else \v{z}\fi{}ek}}, \bibinfo
  {author} {\bibfnamefont {D.}~\bibnamefont {Paar}}, \bibinfo {author}
  {\bibfnamefont {V.}~\bibnamefont {Hinkov}}, \bibinfo {author} {\bibfnamefont
  {M.}~\bibnamefont {Raichle}}, \bibinfo {author} {\bibfnamefont
  {D.}~\bibnamefont {Haug}}, \bibinfo {author} {\bibfnamefont {B.}~\bibnamefont
  {Keimer}}, \bibinfo {author} {\bibfnamefont {N.}~\bibnamefont {Bari\ifmmode
  \check{s}\else \v{s}\fi{}i\ifmmode~\acute{c}\else \'{c}\fi{}}}, \ and\
  \bibinfo {author} {\bibfnamefont {A.}~\bibnamefont {Dul\ifmmode
  \check{c}\else \v{c}\fi{}i\ifmmode~\acute{c}\else \'{c}\fi{}}},\ }\href@noop
  {} {\bibfield  {journal} {\bibinfo  {journal} {Phys. Rev. B}\ }\textbf
  {\bibinfo {volume} {83}},\ \bibinfo {pages} {144508} (\bibinfo {year}
  {2011})}\BibitemShut {NoStop}%
\bibitem [{\citenamefont {Yu}\ \emph {et~al.}(2013)\citenamefont {Yu},
  \citenamefont {Xia}, \citenamefont {Bari\v{s}i\'c}, \citenamefont {He},
  \citenamefont {Kaneko}, \citenamefont {Sasagawa}, \citenamefont {Li},
  \citenamefont {Zhao}, \citenamefont {Shekhter},\ and\ \citenamefont
  {Greven}}]{yu12}%
  \BibitemOpen
  \bibfield  {author} {\bibinfo {author} {\bibfnamefont {G.}~\bibnamefont
  {Yu}}, \bibinfo {author} {\bibfnamefont {D.~D.}\ \bibnamefont {Xia}},
  \bibinfo {author} {\bibfnamefont {N.}~\bibnamefont {Bari\v{s}i\'c}}, \bibinfo
  {author} {\bibfnamefont {R.~H.}\ \bibnamefont {He}}, \bibinfo {author}
  {\bibfnamefont {N.}~\bibnamefont {Kaneko}}, \bibinfo {author} {\bibfnamefont
  {T.}~\bibnamefont {Sasagawa}}, \bibinfo {author} {\bibfnamefont
  {Y.}~\bibnamefont {Li}}, \bibinfo {author} {\bibfnamefont {X.}~\bibnamefont
  {Zhao}}, \bibinfo {author} {\bibfnamefont {A.}~\bibnamefont {Shekhter}}, \
  and\ \bibinfo {author} {\bibfnamefont {M.}~\bibnamefont {Greven}},\
  }\href@noop {} {\bibfield  {journal} {\bibinfo  {journal} {arxiv:1210.6942}\
  } (\bibinfo {year} {2013})}\BibitemShut {NoStop}%
\bibitem [{\citenamefont {Rullier-Albenque}\ \emph {et~al.}(2007)\citenamefont
  {Rullier-Albenque}, \citenamefont {Alloul}, \citenamefont {Proust},
  \citenamefont {Lejay}, \citenamefont {Forget},\ and\ \citenamefont
  {Colson}}]{albenque07}%
  \BibitemOpen
  \bibfield  {author} {\bibinfo {author} {\bibfnamefont {F.}~\bibnamefont
  {Rullier-Albenque}}, \bibinfo {author} {\bibfnamefont {H.}~\bibnamefont
  {Alloul}}, \bibinfo {author} {\bibfnamefont {C.}~\bibnamefont {Proust}},
  \bibinfo {author} {\bibfnamefont {P.}~\bibnamefont {Lejay}}, \bibinfo
  {author} {\bibfnamefont {A.}~\bibnamefont {Forget}}, \ and\ \bibinfo {author}
  {\bibfnamefont {D.}~\bibnamefont {Colson}},\ }\href@noop {} {\bibfield
  {journal} {\bibinfo  {journal} {Phys. Rev. Lett.}\ }\textbf {\bibinfo
  {volume} {99}},\ \bibinfo {pages} {027003} (\bibinfo {year}
  {2007})}\BibitemShut {NoStop}%
\bibitem [{\citenamefont {Bari\v{s}i\'c}\ \emph
  {et~al.}(2013{\natexlab{a}})\citenamefont {Bari\v{s}i\'c}, \citenamefont
  {Chan}, \citenamefont {Li}, \citenamefont {Yu}, \citenamefont {Zhao},
  \citenamefont {Dressel}, \citenamefont {Smontara},\ and\ \citenamefont
  {Greven}}]{barisic13}%
  \BibitemOpen
  \bibfield  {author} {\bibinfo {author} {\bibfnamefont {N.}~\bibnamefont
  {Bari\v{s}i\'c}}, \bibinfo {author} {\bibfnamefont {M.~K.}\ \bibnamefont
  {Chan}}, \bibinfo {author} {\bibfnamefont {Y.}~\bibnamefont {Li}}, \bibinfo
  {author} {\bibfnamefont {G.}~\bibnamefont {Yu}}, \bibinfo {author}
  {\bibfnamefont {X.}~\bibnamefont {Zhao}}, \bibinfo {author} {\bibfnamefont
  {M.}~\bibnamefont {Dressel}}, \bibinfo {author} {\bibfnamefont
  {A.}~\bibnamefont {Smontara}}, \ and\ \bibinfo {author} {\bibfnamefont
  {M.}~\bibnamefont {Greven}},\ }\href@noop {} {\bibfield  {journal} {\bibinfo
  {journal} {Proc. Natl. Acad. Sci.}\ }\textbf {\bibinfo {volume} {110}},\
  \bibinfo {pages} {12235} (\bibinfo {year} {2013}{\natexlab{a}})}\BibitemShut
  {NoStop}%
\bibitem [{\citenamefont {Mirzaei}\ \emph {et~al.}(2013)\citenamefont
  {Mirzaei}, \citenamefont {Stricker}, \citenamefont {Hancock}, \citenamefont
  {Berthod}, \citenamefont {Georges}, \citenamefont {van Heumen}, \citenamefont
  {Chan}, \citenamefont {Zhao}, \citenamefont {Li}, \citenamefont {Greven},
  \citenamefont {Bari\v{s}i\'c},\ and\ \citenamefont {van~der
  Marel}}]{mirzaei13}%
  \BibitemOpen
  \bibfield  {author} {\bibinfo {author} {\bibfnamefont {S.}~\bibnamefont
  {Mirzaei}}, \bibinfo {author} {\bibfnamefont {D.}~\bibnamefont {Stricker}},
  \bibinfo {author} {\bibfnamefont {J.}~\bibnamefont {Hancock}}, \bibinfo
  {author} {\bibfnamefont {C.}~\bibnamefont {Berthod}}, \bibinfo {author}
  {\bibfnamefont {A.}~\bibnamefont {Georges}}, \bibinfo {author} {\bibfnamefont
  {E.}~\bibnamefont {van Heumen}}, \bibinfo {author} {\bibfnamefont {M.~K.}\
  \bibnamefont {Chan}}, \bibinfo {author} {\bibfnamefont {X.}~\bibnamefont
  {Zhao}}, \bibinfo {author} {\bibfnamefont {Y.}~\bibnamefont {Li}}, \bibinfo
  {author} {\bibfnamefont {M.}~\bibnamefont {Greven}}, \bibinfo {author}
  {\bibfnamefont {N.}~\bibnamefont {Bari\v{s}i\'c}}, \ and\ \bibinfo {author}
  {\bibfnamefont {D.}~\bibnamefont {van~der Marel}},\ }\href@noop {} {\bibfield
   {journal} {\bibinfo  {journal} {Proc. Natl. Acad. Sci.}\ }\textbf {\bibinfo
  {volume} {110}},\ \bibinfo {pages} {5774} (\bibinfo {year}
  {2013})}\BibitemShut {NoStop}%
\bibitem [{\citenamefont {Harris}\ \emph {et~al.}(1995)\citenamefont {Harris},
  \citenamefont {Yan}, \citenamefont {Matl}, \citenamefont {Ong}, \citenamefont
  {Anderson}, \citenamefont {Kimura},\ and\ \citenamefont
  {Kitazawa}}]{harris95}%
  \BibitemOpen
  \bibfield  {author} {\bibinfo {author} {\bibfnamefont {M.}~\bibnamefont
  {Harris}}, \bibinfo {author} {\bibfnamefont {Y.}~\bibnamefont {Yan}},
  \bibinfo {author} {\bibfnamefont {P.}~\bibnamefont {Matl}}, \bibinfo {author}
  {\bibfnamefont {N.~P.}\ \bibnamefont {Ong}}, \bibinfo {author} {\bibfnamefont
  {P.~W.}\ \bibnamefont {Anderson}}, \bibinfo {author} {\bibfnamefont
  {T.}~\bibnamefont {Kimura}}, \ and\ \bibinfo {author} {\bibfnamefont
  {K.}~\bibnamefont {Kitazawa}},\ }\href@noop {} {\bibfield  {journal}
  {\bibinfo  {journal} {Phys. Rev. Lett.}\ }\textbf {\bibinfo {volume} {75}},\
  \bibinfo {pages} {1391} (\bibinfo {year} {1995})}\BibitemShut {NoStop}%
\bibitem [{\citenamefont {Semba}\ and\ \citenamefont
  {Matsuda}(1997)}]{semba97}%
  \BibitemOpen
  \bibfield  {author} {\bibinfo {author} {\bibfnamefont {K.}~\bibnamefont
  {Semba}}\ and\ \bibinfo {author} {\bibfnamefont {A.}~\bibnamefont
  {Matsuda}},\ }\href@noop {} {\bibfield  {journal} {\bibinfo  {journal} {Phys.
  Rev. B}\ }\textbf {\bibinfo {volume} {55}},\ \bibinfo {pages} {11103}
  (\bibinfo {year} {1997})}\BibitemShut {NoStop}%
\bibitem [{\citenamefont {Kimura}\ \emph {et~al.}(1996)\citenamefont {Kimura},
  \citenamefont {Miyasaka}, \citenamefont {Takagi}, \citenamefont {Tamasaku},
  \citenamefont {Eisaki}, \citenamefont {Uchida}, \citenamefont {Kitazawa},
  \citenamefont {Hiroi}, \citenamefont {Sera},\ and\ \citenamefont
  {Kobayashi}}]{kimura96}%
  \BibitemOpen
  \bibfield  {author} {\bibinfo {author} {\bibfnamefont {T.}~\bibnamefont
  {Kimura}}, \bibinfo {author} {\bibfnamefont {S.}~\bibnamefont {Miyasaka}},
  \bibinfo {author} {\bibfnamefont {H.}~\bibnamefont {Takagi}}, \bibinfo
  {author} {\bibfnamefont {K.}~\bibnamefont {Tamasaku}}, \bibinfo {author}
  {\bibfnamefont {H.}~\bibnamefont {Eisaki}}, \bibinfo {author} {\bibfnamefont
  {S.}~\bibnamefont {Uchida}}, \bibinfo {author} {\bibfnamefont
  {K.}~\bibnamefont {Kitazawa}}, \bibinfo {author} {\bibfnamefont
  {M.}~\bibnamefont {Hiroi}}, \bibinfo {author} {\bibfnamefont
  {M.}~\bibnamefont {Sera}}, \ and\ \bibinfo {author} {\bibfnamefont
  {N.}~\bibnamefont {Kobayashi}},\ }\href@noop {} {\bibfield  {journal}
  {\bibinfo  {journal} {Phys. Rev. B}\ }\textbf {\bibinfo {volume} {53}},\
  \bibinfo {pages} {8733} (\bibinfo {year} {1996})}\BibitemShut {NoStop}%
\bibitem [{\citenamefont {Eisaki}\ \emph {et~al.}(2004)\citenamefont {Eisaki},
  \citenamefont {Kaneko}, \citenamefont {Feng}, \citenamefont {Damascelli},
  \citenamefont {Mang}, \citenamefont {Shen}, \citenamefont {Shen},\ and\
  \citenamefont {Greven}}]{eisaki04}%
  \BibitemOpen
  \bibfield  {author} {\bibinfo {author} {\bibfnamefont {H.}~\bibnamefont
  {Eisaki}}, \bibinfo {author} {\bibfnamefont {N.}~\bibnamefont {Kaneko}},
  \bibinfo {author} {\bibfnamefont {D.~L.}\ \bibnamefont {Feng}}, \bibinfo
  {author} {\bibfnamefont {A.}~\bibnamefont {Damascelli}}, \bibinfo {author}
  {\bibfnamefont {P.~K.}\ \bibnamefont {Mang}}, \bibinfo {author}
  {\bibfnamefont {K.~M.}\ \bibnamefont {Shen}}, \bibinfo {author}
  {\bibfnamefont {Z.-X.}\ \bibnamefont {Shen}}, \ and\ \bibinfo {author}
  {\bibfnamefont {M.}~\bibnamefont {Greven}},\ }\href@noop {} {\bibfield
  {journal} {\bibinfo  {journal} {Phys. Rev. B}\ }\textbf {\bibinfo {volume}
  {69}},\ \bibinfo {pages} {064512} (\bibinfo {year} {2004})}\BibitemShut
  {NoStop}%
\bibitem [{\citenamefont {Putilin}\ \emph {et~al.}(1993)\citenamefont
  {Putilin}, \citenamefont {Antipov}, \citenamefont {Chmaissem},\ and\
  \citenamefont {Marezio}}]{putilin93}%
  \BibitemOpen
  \bibfield  {author} {\bibinfo {author} {\bibfnamefont {S.~N.}\ \bibnamefont
  {Putilin}}, \bibinfo {author} {\bibfnamefont {E.~V.}\ \bibnamefont
  {Antipov}}, \bibinfo {author} {\bibfnamefont {O.}~\bibnamefont {Chmaissem}},
  \ and\ \bibinfo {author} {\bibfnamefont {M.}~\bibnamefont {Marezio}},\
  }\href@noop {} {\bibfield  {journal} {\bibinfo  {journal} {Nature (London)}\
  }\textbf {\bibinfo {volume} {362}},\ \bibinfo {pages} {226} (\bibinfo {year}
  {1993})}\BibitemShut {NoStop}%
\bibitem [{\citenamefont {Zhao}\ \emph {et~al.}(2006)\citenamefont {Zhao},
  \citenamefont {Yu}, \citenamefont {Cho}, \citenamefont {Chabot-Couture},
  \citenamefont {Bari\v{s}i\'c}, \citenamefont {Bourges}, \citenamefont
  {Kaneko}, \citenamefont {Li}, \citenamefont {Lu}, \citenamefont {Motoyama},
  \citenamefont {Vajk},\ and\ \citenamefont {Greven}}]{zhao06}%
  \BibitemOpen
  \bibfield  {author} {\bibinfo {author} {\bibfnamefont {X.}~\bibnamefont
  {Zhao}}, \bibinfo {author} {\bibfnamefont {G.}~\bibnamefont {Yu}}, \bibinfo
  {author} {\bibfnamefont {Y.-C.}\ \bibnamefont {Cho}}, \bibinfo {author}
  {\bibfnamefont {G.}~\bibnamefont {Chabot-Couture}}, \bibinfo {author}
  {\bibfnamefont {N.}~\bibnamefont {Bari\v{s}i\'c}}, \bibinfo {author}
  {\bibfnamefont {P.}~\bibnamefont {Bourges}}, \bibinfo {author} {\bibfnamefont
  {N.}~\bibnamefont {Kaneko}}, \bibinfo {author} {\bibfnamefont
  {Y.}~\bibnamefont {Li}}, \bibinfo {author} {\bibfnamefont {L.}~\bibnamefont
  {Lu}}, \bibinfo {author} {\bibfnamefont {E.~M.}\ \bibnamefont {Motoyama}},
  \bibinfo {author} {\bibfnamefont {O.~P.}\ \bibnamefont {Vajk}}, \ and\
  \bibinfo {author} {\bibfnamefont {M.}~\bibnamefont {Greven}},\ }\href@noop {}
  {\bibfield  {journal} {\bibinfo  {journal} {Adv. Mater.}\ }\textbf {\bibinfo
  {volume} {18}},\ \bibinfo {pages} {3243} (\bibinfo {year}
  {2006})}\BibitemShut {NoStop}%
\bibitem [{\citenamefont {Bari\v{s}i\'c}\ \emph {et~al.}(2008)\citenamefont
  {Bari\v{s}i\'c}, \citenamefont {Li}, \citenamefont {Zhao}, \citenamefont
  {Cho}, \citenamefont {Chabot-Couture}, \citenamefont {Yu},\ and\
  \citenamefont {Greven}}]{barisic08}%
  \BibitemOpen
  \bibfield  {author} {\bibinfo {author} {\bibfnamefont {N.}~\bibnamefont
  {Bari\v{s}i\'c}}, \bibinfo {author} {\bibfnamefont {Y.}~\bibnamefont {Li}},
  \bibinfo {author} {\bibfnamefont {X.}~\bibnamefont {Zhao}}, \bibinfo {author}
  {\bibfnamefont {Y.-C.}\ \bibnamefont {Cho}}, \bibinfo {author} {\bibfnamefont
  {G.}~\bibnamefont {Chabot-Couture}}, \bibinfo {author} {\bibfnamefont
  {G.}~\bibnamefont {Yu}}, \ and\ \bibinfo {author} {\bibfnamefont
  {M.}~\bibnamefont {Greven}},\ }\href@noop {} {\bibfield  {journal} {\bibinfo
  {journal} {Phys. Rev. B}\ }\textbf {\bibinfo {volume} {78}},\ \bibinfo
  {pages} {054518} (\bibinfo {year} {2008})}\BibitemShut {NoStop}%
\bibitem [{\citenamefont {Li}\ \emph {et~al.}(2011)\citenamefont {Li},
  \citenamefont {Egetenmeyer}, \citenamefont {Gavilano}, \citenamefont
  {Bari\ifmmode \check{s}\else \v{s}\fi{}i\ifmmode~\acute{c}\else \'{c}\fi{}},\
  and\ \citenamefont {Greven}}]{li11}%
  \BibitemOpen
  \bibfield  {author} {\bibinfo {author} {\bibfnamefont {Y.}~\bibnamefont
  {Li}}, \bibinfo {author} {\bibfnamefont {N.}~\bibnamefont {Egetenmeyer}},
  \bibinfo {author} {\bibfnamefont {J.~L.}\ \bibnamefont {Gavilano}}, \bibinfo
  {author} {\bibfnamefont {N.}~\bibnamefont {Bari\ifmmode \check{s}\else
  \v{s}\fi{}i\ifmmode~\acute{c}\else \'{c}\fi{}}}, \ and\ \bibinfo {author}
  {\bibfnamefont {M.}~\bibnamefont {Greven}},\ }\href {\doibase
  10.1103/PhysRevB.83.054507} {\bibfield  {journal} {\bibinfo  {journal} {Phys.
  Rev. B}\ }\textbf {\bibinfo {volume} {83}},\ \bibinfo {pages} {054507}
  (\bibinfo {year} {2011})}\BibitemShut {NoStop}%
\bibitem [{\citenamefont {Bari\v{s}i\'c}\ \emph
  {et~al.}(2013{\natexlab{b}})\citenamefont {Bari\v{s}i\'c}, \citenamefont
  {Badoux}, \citenamefont {Chan}, \citenamefont {Dorow}, \citenamefont {Tabis},
  \citenamefont {Vignolle}, \citenamefont {Yu}, \citenamefont {B\'{e}ard},
  \citenamefont {Zhao}, \citenamefont {Proust},\ and\ \citenamefont
  {Greven}}]{barisic13b}%
  \BibitemOpen
  \bibfield  {author} {\bibinfo {author} {\bibfnamefont {N.}~\bibnamefont
  {Bari\v{s}i\'c}}, \bibinfo {author} {\bibfnamefont {S.}~\bibnamefont
  {Badoux}}, \bibinfo {author} {\bibfnamefont {M.~K.}\ \bibnamefont {Chan}},
  \bibinfo {author} {\bibfnamefont {C.}~\bibnamefont {Dorow}}, \bibinfo
  {author} {\bibfnamefont {W.}~\bibnamefont {Tabis}}, \bibinfo {author}
  {\bibfnamefont {B.}~\bibnamefont {Vignolle}}, \bibinfo {author}
  {\bibfnamefont {G.}~\bibnamefont {Yu}}, \bibinfo {author} {\bibfnamefont
  {J.}~\bibnamefont {B\'{e}ard}}, \bibinfo {author} {\bibfnamefont
  {X.}~\bibnamefont {Zhao}}, \bibinfo {author} {\bibnamefont {Proust}}, \ and\
  \bibinfo {author} {\bibfnamefont {M.}~\bibnamefont {Greven}},\ }\href@noop {}
  {\bibfield  {journal} {\bibinfo  {journal} {Nature Phys.}\ }\textbf {\bibinfo
  {volume} {9}},\ \bibinfo {pages} {761} (\bibinfo {year}
  {2013}{\natexlab{b}})}\BibitemShut {NoStop}%
\bibitem [{\citenamefont {Doiron-Leyraud}\ \emph {et~al.}(2013)\citenamefont
  {Doiron-Leyraud}, \citenamefont {Lepault}, \citenamefont {Cyr-Choini\'{e}re},
  \citenamefont {Vignolle}, \citenamefont {Grissonnanche}, \citenamefont
  {Lalibert\'{e}}, \citenamefont {Chang}, \citenamefont {Bari\v{s}i\'c},
  \citenamefont {Chan}, \citenamefont {Ji}, \citenamefont {Zhao}, \citenamefont
  {Li}, \citenamefont {Greven}, \citenamefont {Proust},\ and\ \citenamefont
  {Taillefer}}]{leyraud13}%
  \BibitemOpen
  \bibfield  {author} {\bibinfo {author} {\bibfnamefont {N.}~\bibnamefont
  {Doiron-Leyraud}}, \bibinfo {author} {\bibfnamefont {S.}~\bibnamefont
  {Lepault}}, \bibinfo {author} {\bibfnamefont {O.}~\bibnamefont
  {Cyr-Choini\'{e}re}}, \bibinfo {author} {\bibfnamefont {B.}~\bibnamefont
  {Vignolle}}, \bibinfo {author} {\bibfnamefont {G.}~\bibnamefont
  {Grissonnanche}}, \bibinfo {author} {\bibfnamefont {F.}~\bibnamefont
  {Lalibert\'{e}}}, \bibinfo {author} {\bibfnamefont {J.}~\bibnamefont
  {Chang}}, \bibinfo {author} {\bibfnamefont {N.}~\bibnamefont
  {Bari\v{s}i\'c}}, \bibinfo {author} {\bibfnamefont {M.~K.}\ \bibnamefont
  {Chan}}, \bibinfo {author} {\bibfnamefont {L.}~\bibnamefont {Ji}}, \bibinfo
  {author} {\bibfnamefont {X.}~\bibnamefont {Zhao}}, \bibinfo {author}
  {\bibfnamefont {Y.}~\bibnamefont {Li}}, \bibinfo {author} {\bibfnamefont
  {M.}~\bibnamefont {Greven}}, \bibinfo {author} {\bibfnamefont
  {C.}~\bibnamefont {Proust}}, \ and\ \bibinfo {author} {\bibfnamefont
  {L.}~\bibnamefont {Taillefer}},\ }\href@noop {} {\bibfield  {journal}
  {\bibinfo  {journal} {Phys. Rev. X}\ }\textbf {\bibinfo {volume} {3}},\
  \bibinfo {pages} {021019} (\bibinfo {year} {2013})}\BibitemShut {NoStop}%
\bibitem [{\citenamefont {Yamamoto}\ \emph {et~al.}(2000)\citenamefont
  {Yamamoto}, \citenamefont {Hu},\ and\ \citenamefont {Tajima}}]{yamamoto00}%
  \BibitemOpen
  \bibfield  {author} {\bibinfo {author} {\bibfnamefont {A.}~\bibnamefont
  {Yamamoto}}, \bibinfo {author} {\bibfnamefont {Q.-Z.}\ \bibnamefont {Hu}}, \
  and\ \bibinfo {author} {\bibfnamefont {S.}~\bibnamefont {Tajima}},\
  }\href@noop {} {\bibfield  {journal} {\bibinfo  {journal} {Phys. Rev. B}\
  }\textbf {\bibinfo {volume} {63}},\ \bibinfo {pages} {024504} (\bibinfo
  {year} {2000})}\BibitemShut {NoStop}%
\bibitem [{nhm()}]{nhmfl}%
  \BibitemOpen
  \href@noop {} {}\bibinfo {note} {We have recently performed additional
  magnetotransport measurements on a sample with the same doping in {\it
  static} fields up to 30 T at the National High Magnetic Field Laboratory,
  Tallahassee, FL, USA. These measurements are consistent with the results
  presented here and further confirm the validity of Kohler's rule from 100 K
  to 230 K in another sample with $T_c = 70 K$}\BibitemShut {NoStop}%
\bibitem [{\citenamefont {Cooper}\ \emph {et~al.}(2009)\citenamefont {Cooper},
  \citenamefont {Wang}, \citenamefont {Vignolle}, \citenamefont {Lipscombe},
  \citenamefont {Hayden}, \citenamefont {Tanabe}, \citenamefont {Adachi},
  \citenamefont {Koike}, \citenamefont {Nohara}, \citenamefont {Takagi},
  \citenamefont {Proust},\ and\ \citenamefont {Hussey}}]{cooper09}%
  \BibitemOpen
  \bibfield  {author} {\bibinfo {author} {\bibfnamefont {R.~A.}\ \bibnamefont
  {Cooper}}, \bibinfo {author} {\bibfnamefont {Y.}~\bibnamefont {Wang}},
  \bibinfo {author} {\bibfnamefont {B.}~\bibnamefont {Vignolle}}, \bibinfo
  {author} {\bibfnamefont {O.~J.}\ \bibnamefont {Lipscombe}}, \bibinfo {author}
  {\bibfnamefont {S.~M.}\ \bibnamefont {Hayden}}, \bibinfo {author}
  {\bibfnamefont {Y.}~\bibnamefont {Tanabe}}, \bibinfo {author} {\bibfnamefont
  {T.}~\bibnamefont {Adachi}}, \bibinfo {author} {\bibfnamefont
  {Y.}~\bibnamefont {Koike}}, \bibinfo {author} {\bibfnamefont
  {M.}~\bibnamefont {Nohara}}, \bibinfo {author} {\bibfnamefont
  {H.}~\bibnamefont {Takagi}}, \bibinfo {author} {\bibfnamefont
  {C.}~\bibnamefont {Proust}}, \ and\ \bibinfo {author} {\bibfnamefont {N.~E.}\
  \bibnamefont {Hussey}},\ }\href@noop {} {\bibfield  {journal} {\bibinfo
  {journal} {Science}\ }\textbf {\bibinfo {volume} {323}},\ \bibinfo {pages}
  {603} (\bibinfo {year} {2009})}\BibitemShut {NoStop}%
\bibitem [{\citenamefont {Kokanovi\'{c}}\ \emph {et~al.}(2006)\citenamefont
  {Kokanovi\'{c}}, \citenamefont {Cooper}, \citenamefont {Naqib},\ and\
  \citenamefont {Islam}}]{koka06}%
  \BibitemOpen
  \bibfield  {author} {\bibinfo {author} {\bibfnamefont {I.}~\bibnamefont
  {Kokanovi\'{c}}}, \bibinfo {author} {\bibfnamefont {J.~R.}\ \bibnamefont
  {Cooper}}, \bibinfo {author} {\bibfnamefont {S.~H.}\ \bibnamefont {Naqib}}, \
  and\ \bibinfo {author} {\bibfnamefont {R.~S.}\ \bibnamefont {Islam}},\
  }\href@noop {} {\bibfield  {journal} {\bibinfo  {journal} {Phys. Rev. B}\
  }\textbf {\bibinfo {volume} {73}},\ \bibinfo {pages} {184509} (\bibinfo
  {year} {2006})}\BibitemShut {NoStop}%
\bibitem [{\citenamefont {Ando}\ and\ \citenamefont {Segawa}(2002)}]{ando02}%
  \BibitemOpen
  \bibfield  {author} {\bibinfo {author} {\bibfnamefont {Y.}~\bibnamefont
  {Ando}}\ and\ \bibinfo {author} {\bibfnamefont {K.}~\bibnamefont {Segawa}},\
  }\href@noop {} {\bibfield  {journal} {\bibinfo  {journal} {Phys. Rev. Lett.}\
  }\textbf {\bibinfo {volume} {88}},\ \bibinfo {pages} {167005} (\bibinfo
  {year} {2002})}\BibitemShut {NoStop}%
\bibitem [{\citenamefont {Liang}\ \emph {et~al.}()\citenamefont {Liang},
  \citenamefont {Bonn},\ and\ \citenamefont {Hardy}}]{liang06}%
  \BibitemOpen
  \bibfield  {author} {\bibinfo {author} {\bibfnamefont {R.}~\bibnamefont
  {Liang}}, \bibinfo {author} {\bibfnamefont {D.~A.}\ \bibnamefont {Bonn}}, \
  and\ \bibinfo {author} {\bibfnamefont {W.~N.}\ \bibnamefont {Hardy}},\
  }\href@noop {} {\bibfield  {journal} {\bibinfo  {journal} {Phys. Rev. B}\
  }\textbf {\bibinfo {volume} {73}},\ \bibinfo {pages} {180505}}\BibitemShut
  {NoStop}%
\bibitem [{\citenamefont {Lee}\ \emph {et~al.}(2005)\citenamefont {Lee},
  \citenamefont {Segawa}, \citenamefont {Li}, \citenamefont {Padilla},
  \citenamefont {Dumm}, \citenamefont {Dordevic}, \citenamefont {Homes},
  \citenamefont {Ando},\ and\ \citenamefont {Basov}}]{lee05}%
  \BibitemOpen
  \bibfield  {author} {\bibinfo {author} {\bibfnamefont {Y.~S.}\ \bibnamefont
  {Lee}}, \bibinfo {author} {\bibfnamefont {K.}~\bibnamefont {Segawa}},
  \bibinfo {author} {\bibfnamefont {Z.~Q.}\ \bibnamefont {Li}}, \bibinfo
  {author} {\bibfnamefont {W.~J.}\ \bibnamefont {Padilla}}, \bibinfo {author}
  {\bibfnamefont {M.}~\bibnamefont {Dumm}}, \bibinfo {author} {\bibfnamefont
  {S.~V.}\ \bibnamefont {Dordevic}}, \bibinfo {author} {\bibfnamefont {C.~C.}\
  \bibnamefont {Homes}}, \bibinfo {author} {\bibfnamefont {Y.}~\bibnamefont
  {Ando}}, \ and\ \bibinfo {author} {\bibfnamefont {D.~N.}\ \bibnamefont
  {Basov}},\ }\href@noop {} {\bibfield  {journal} {\bibinfo  {journal} {Phys.
  Rev. B}\ }\textbf {\bibinfo {volume} {72(5)}},\ \bibinfo {pages} {054529}
  (\bibinfo {year} {2005})}\BibitemShut {NoStop}%
\bibitem [{\citenamefont {Boebinger}\ \emph {et~al.}(1996)\citenamefont
  {Boebinger}, \citenamefont {Ando}, \citenamefont {Passner}, \citenamefont
  {Kimura}, \citenamefont {Okuya}, \citenamefont {Shimoyama}, \citenamefont
  {Kishio}, \citenamefont {Tamasaku}, \citenamefont {Ichikawa},\ and\
  \citenamefont {Uchida}}]{boebinger96}%
  \BibitemOpen
  \bibfield  {author} {\bibinfo {author} {\bibfnamefont {G.~S.}\ \bibnamefont
  {Boebinger}}, \bibinfo {author} {\bibfnamefont {Y.}~\bibnamefont {Ando}},
  \bibinfo {author} {\bibfnamefont {A.}~\bibnamefont {Passner}}, \bibinfo
  {author} {\bibfnamefont {T.}~\bibnamefont {Kimura}}, \bibinfo {author}
  {\bibfnamefont {M.}~\bibnamefont {Okuya}}, \bibinfo {author} {\bibfnamefont
  {J.}~\bibnamefont {Shimoyama}}, \bibinfo {author} {\bibfnamefont
  {K.}~\bibnamefont {Kishio}}, \bibinfo {author} {\bibfnamefont
  {K.}~\bibnamefont {Tamasaku}}, \bibinfo {author} {\bibfnamefont
  {N.}~\bibnamefont {Ichikawa}}, \ and\ \bibinfo {author} {\bibfnamefont
  {S.}~\bibnamefont {Uchida}},\ }\href@noop {} {\bibfield  {journal} {\bibinfo
  {journal} {Phys. Rev. Lett.}\ }\textbf {\bibinfo {volume} {77}},\ \bibinfo
  {pages} {5417} (\bibinfo {year} {1996})}\BibitemShut {NoStop}%
\bibitem [{\citenamefont {Doiron-Leyraud}\ \emph {et~al.}(2007)\citenamefont
  {Doiron-Leyraud}, \citenamefont {Proust}, \citenamefont {LeBoeuf},
  \citenamefont {Levallois}, \citenamefont {Bonnemaison}, \citenamefont
  {Liang}, \citenamefont {Bonn}, \citenamefont {Hardy},\ and\ \citenamefont
  {Taillefer}}]{leyraud07}%
  \BibitemOpen
  \bibfield  {author} {\bibinfo {author} {\bibfnamefont {N.}~\bibnamefont
  {Doiron-Leyraud}}, \bibinfo {author} {\bibfnamefont {C.}~\bibnamefont
  {Proust}}, \bibinfo {author} {\bibfnamefont {D.}~\bibnamefont {LeBoeuf}},
  \bibinfo {author} {\bibfnamefont {J.}~\bibnamefont {Levallois}}, \bibinfo
  {author} {\bibfnamefont {J.}~\bibnamefont {Bonnemaison}}, \bibinfo {author}
  {\bibfnamefont {R.}~\bibnamefont {Liang}}, \bibinfo {author} {\bibfnamefont
  {D.~A.}\ \bibnamefont {Bonn}}, \bibinfo {author} {\bibfnamefont {W.~N.}\
  \bibnamefont {Hardy}}, \ and\ \bibinfo {author} {\bibfnamefont
  {L.}~\bibnamefont {Taillefer}},\ }\href@noop {} {\bibfield  {journal}
  {\bibinfo  {journal} {Nature}\ }\textbf {\bibinfo {volume} {447}},\ \bibinfo
  {pages} {565} (\bibinfo {year} {2007})}\BibitemShut {NoStop}%
\bibitem [{\citenamefont {LeBoeuf}\ \emph {et~al.}(2011)\citenamefont
  {LeBoeuf}, \citenamefont {Doiron-Leyraud}, \citenamefont {Vignolle},
  \citenamefont {Sutherland}, \citenamefont {Ramshaw}, \citenamefont
  {Levallois}, \citenamefont {Daou}, \citenamefont {Lalibert\'{e}},
  \citenamefont {Cyr-Choini\'{e}re}, \citenamefont {Chang}, \citenamefont {Jo},
  \citenamefont {Balicas}, \citenamefont {Liang}, \citenamefont {Bonn},
  \citenamefont {Hardy}, \citenamefont {Proust},\ and\ \citenamefont
  {Taillefer}}]{leboeuf11}%
  \BibitemOpen
  \bibfield  {author} {\bibinfo {author} {\bibfnamefont {D.}~\bibnamefont
  {LeBoeuf}}, \bibinfo {author} {\bibfnamefont {N.}~\bibnamefont
  {Doiron-Leyraud}}, \bibinfo {author} {\bibfnamefont {B.}~\bibnamefont
  {Vignolle}}, \bibinfo {author} {\bibfnamefont {M.}~\bibnamefont
  {Sutherland}}, \bibinfo {author} {\bibfnamefont {B.~J.}\ \bibnamefont
  {Ramshaw}}, \bibinfo {author} {\bibfnamefont {J.}~\bibnamefont {Levallois}},
  \bibinfo {author} {\bibfnamefont {R.}~\bibnamefont {Daou}}, \bibinfo {author}
  {\bibfnamefont {F.}~\bibnamefont {Lalibert\'{e}}}, \bibinfo {author}
  {\bibfnamefont {O.}~\bibnamefont {Cyr-Choini\'{e}re}}, \bibinfo {author}
  {\bibfnamefont {J.}~\bibnamefont {Chang}}, \bibinfo {author} {\bibfnamefont
  {Y.~J.}\ \bibnamefont {Jo}}, \bibinfo {author} {\bibfnamefont
  {L.}~\bibnamefont {Balicas}}, \bibinfo {author} {\bibfnamefont
  {R.}~\bibnamefont {Liang}}, \bibinfo {author} {\bibfnamefont {D.~A.}\
  \bibnamefont {Bonn}}, \bibinfo {author} {\bibfnamefont {W.~N.}\ \bibnamefont
  {Hardy}}, \bibinfo {author} {\bibfnamefont {C.}~\bibnamefont {Proust}}, \
  and\ \bibinfo {author} {\bibfnamefont {L.}~\bibnamefont {Taillefer}},\
  }\href@noop {} {\bibfield  {journal} {\bibinfo  {journal} {Phys. Rev. B}\
  }\textbf {\bibinfo {volume} {83}},\ \bibinfo {pages} {054506} (\bibinfo
  {year} {2011})}\BibitemShut {NoStop}%
\bibitem [{\citenamefont {Wang}\ \emph {et~al.}(2006)\citenamefont {Wang},
  \citenamefont {Li},\ and\ \citenamefont {Ong}}]{wang06}%
  \BibitemOpen
  \bibfield  {author} {\bibinfo {author} {\bibfnamefont {Y.}~\bibnamefont
  {Wang}}, \bibinfo {author} {\bibfnamefont {L.}~\bibnamefont {Li}}, \ and\
  \bibinfo {author} {\bibfnamefont {N.}~\bibnamefont {Ong}},\ }\href@noop {}
  {\bibfield  {journal} {\bibinfo  {journal} {Phys. Rev. B}\ }\textbf {\bibinfo
  {volume} {73}},\ \bibinfo {pages} {024510} (\bibinfo {year}
  {2006})}\BibitemShut {NoStop}%
\bibitem [{\citenamefont {Cyr-Choini\'{e}re}\ \emph {et~al.}(2009)\citenamefont
  {Cyr-Choini\'{e}re}, \citenamefont {Daou}, \citenamefont {Laliberte},
  \citenamefont {LeBoeuf}, \citenamefont {Doiron-Leyraud}, \citenamefont
  {Chang}, \citenamefont {Yan}, \citenamefont {Cheng}, \citenamefont {Zhou},
  \citenamefont {Goodenough}, \citenamefont {Pyon}, \citenamefont {Takayama},
  \citenamefont {Takagi}, \citenamefont {Tanaka},\ and\ \citenamefont
  {Taillefer}}]{cyr09}%
  \BibitemOpen
  \bibfield  {author} {\bibinfo {author} {\bibfnamefont {O.}~\bibnamefont
  {Cyr-Choini\'{e}re}}, \bibinfo {author} {\bibfnamefont {R.}~\bibnamefont
  {Daou}}, \bibinfo {author} {\bibfnamefont {F.}~\bibnamefont {Laliberte}},
  \bibinfo {author} {\bibfnamefont {D.}~\bibnamefont {LeBoeuf}}, \bibinfo
  {author} {\bibfnamefont {N.}~\bibnamefont {Doiron-Leyraud}}, \bibinfo
  {author} {\bibfnamefont {J.}~\bibnamefont {Chang}}, \bibinfo {author}
  {\bibfnamefont {J.-Q.}\ \bibnamefont {Yan}}, \bibinfo {author} {\bibfnamefont
  {J.-G.}\ \bibnamefont {Cheng}}, \bibinfo {author} {\bibfnamefont {J.-S.}\
  \bibnamefont {Zhou}}, \bibinfo {author} {\bibfnamefont {J.~B.}\ \bibnamefont
  {Goodenough}}, \bibinfo {author} {\bibfnamefont {S.}~\bibnamefont {Pyon}},
  \bibinfo {author} {\bibfnamefont {T.}~\bibnamefont {Takayama}}, \bibinfo
  {author} {\bibfnamefont {H.}~\bibnamefont {Takagi}}, \bibinfo {author}
  {\bibfnamefont {Y.}~\bibnamefont {Tanaka}}, \ and\ \bibinfo {author}
  {\bibfnamefont {L.}~\bibnamefont {Taillefer}},\ }\href@noop {} {\bibfield
  {journal} {\bibinfo  {journal} {Nature}\ }\textbf {\bibinfo {volume} {458}},\
  \bibinfo {pages} {743} (\bibinfo {year} {2009})}\BibitemShut {NoStop}%
\bibitem [{\citenamefont {Ando}\ \emph
  {et~al.}(2004{\natexlab{b}})\citenamefont {Ando}, \citenamefont {Komiya},
  \citenamefont {Segawa}, \citenamefont {Ono},\ and\ \citenamefont
  {Kurita}}]{ando04b}%
  \BibitemOpen
  \bibfield  {author} {\bibinfo {author} {\bibfnamefont {Y.}~\bibnamefont
  {Ando}}, \bibinfo {author} {\bibfnamefont {S.}~\bibnamefont {Komiya}},
  \bibinfo {author} {\bibfnamefont {K.}~\bibnamefont {Segawa}}, \bibinfo
  {author} {\bibfnamefont {S.}~\bibnamefont {Ono}}, \ and\ \bibinfo {author}
  {\bibfnamefont {Y.}~\bibnamefont {Kurita}},\ }\href@noop {} {\bibfield
  {journal} {\bibinfo  {journal} {Phys. Rev. Lett.}\ }\textbf {\bibinfo
  {volume} {93}},\ \bibinfo {pages} {267001} (\bibinfo {year}
  {2004}{\natexlab{b}})}\BibitemShut {NoStop}%
\bibitem [{\citenamefont {Bollinger}\ \emph {et~al.}(2011)\citenamefont
  {Bollinger}, \citenamefont {Dubuis}, \citenamefont {Yoon}, \citenamefont
  {Pavuna}, \citenamefont {Misewich},\ and\ \citenamefont
  {Bo\v{z}ovi\'c}}]{bollinger11}%
  \BibitemOpen
  \bibfield  {author} {\bibinfo {author} {\bibfnamefont {A.~T.}\ \bibnamefont
  {Bollinger}}, \bibinfo {author} {\bibfnamefont {G.}~\bibnamefont {Dubuis}},
  \bibinfo {author} {\bibfnamefont {J.}~\bibnamefont {Yoon}}, \bibinfo {author}
  {\bibfnamefont {D.}~\bibnamefont {Pavuna}}, \bibinfo {author} {\bibfnamefont
  {J.}~\bibnamefont {Misewich}}, \ and\ \bibinfo {author} {\bibfnamefont
  {I.}~\bibnamefont {Bo\v{z}ovi\'c}},\ }\href@noop {} {\bibfield  {journal}
  {\bibinfo  {journal} {Nature}\ }\textbf {\bibinfo {volume} {472}},\ \bibinfo
  {pages} {458} (\bibinfo {year} {2011})}\BibitemShut {NoStop}%
\bibitem [{\citenamefont {Walker}\ \emph {et~al.}(1995)\citenamefont {Walker},
  \citenamefont {Mackenzie},\ and\ \citenamefont {Cooper}}]{walker95}%
  \BibitemOpen
  \bibfield  {author} {\bibinfo {author} {\bibfnamefont {D.~J.~C.}\
  \bibnamefont {Walker}}, \bibinfo {author} {\bibfnamefont {A.~P.}\
  \bibnamefont {Mackenzie}}, \ and\ \bibinfo {author} {\bibfnamefont {J.~R.}\
  \bibnamefont {Cooper}},\ }\href@noop {} {\bibfield  {journal} {\bibinfo
  {journal} {Phys. Rev. B}\ }\textbf {\bibinfo {volume} {51}},\ \bibinfo
  {pages} {15653} (\bibinfo {year} {1995})}\BibitemShut {NoStop}%
\bibitem [{\citenamefont {Leng}\ \emph {et~al.}(2011)\citenamefont {Leng},
  \citenamefont {Garcia-Barriocanal}, \citenamefont {Bose}, \citenamefont
  {Lee},\ and\ \citenamefont {Goldman}}]{leng11}%
  \BibitemOpen
  \bibfield  {author} {\bibinfo {author} {\bibfnamefont {X.}~\bibnamefont
  {Leng}}, \bibinfo {author} {\bibfnamefont {J.}~\bibnamefont
  {Garcia-Barriocanal}}, \bibinfo {author} {\bibfnamefont {S.}~\bibnamefont
  {Bose}}, \bibinfo {author} {\bibfnamefont {Y.}~\bibnamefont {Lee}}, \ and\
  \bibinfo {author} {\bibfnamefont {A.~M.}\ \bibnamefont {Goldman}},\
  }\href@noop {} {\bibfield  {journal} {\bibinfo  {journal} {Phys. Rev. Lett.}\
  }\textbf {\bibinfo {volume} {107}},\ \bibinfo {pages} {027001} (\bibinfo
  {year} {2011})}\BibitemShut {NoStop}%
\bibitem [{\citenamefont {Mandrus}\ \emph {et~al.}(1991)\citenamefont
  {Mandrus}, \citenamefont {Forro}, \citenamefont {Kendziora},\ and\
  \citenamefont {Mihaly}}]{mandrus91}%
  \BibitemOpen
  \bibfield  {author} {\bibinfo {author} {\bibfnamefont {D.}~\bibnamefont
  {Mandrus}}, \bibinfo {author} {\bibfnamefont {L.}~\bibnamefont {Forro}},
  \bibinfo {author} {\bibfnamefont {C.}~\bibnamefont {Kendziora}}, \ and\
  \bibinfo {author} {\bibfnamefont {L.}~\bibnamefont {Mihaly}},\ }\href@noop {}
  {\bibfield  {journal} {\bibinfo  {journal} {Phys. Rev. B}\ }\textbf {\bibinfo
  {volume} {44}},\ \bibinfo {pages} {2418} (\bibinfo {year}
  {1991})}\BibitemShut {NoStop}%
\bibitem [{\citenamefont {Yamada}\ \emph {et~al.}(1998)\citenamefont {Yamada},
  \citenamefont {Lee}, \citenamefont {Kurahashi}, \citenamefont {Wada},
  \citenamefont {Wakimoto}, \citenamefont {Ueki}, \citenamefont {Kimura},
  \citenamefont {Endoh}, \citenamefont {Hosoya}, \citenamefont {Shirane},
  \citenamefont {Birgeneau}, \citenamefont {Greven}, \citenamefont {Kastner},\
  and\ \citenamefont {Kim}}]{yamada98}%
  \BibitemOpen
  \bibfield  {author} {\bibinfo {author} {\bibfnamefont {K.}~\bibnamefont
  {Yamada}}, \bibinfo {author} {\bibfnamefont {C.~H.}\ \bibnamefont {Lee}},
  \bibinfo {author} {\bibfnamefont {K.}~\bibnamefont {Kurahashi}}, \bibinfo
  {author} {\bibfnamefont {J.}~\bibnamefont {Wada}}, \bibinfo {author}
  {\bibfnamefont {S.}~\bibnamefont {Wakimoto}}, \bibinfo {author}
  {\bibfnamefont {S.}~\bibnamefont {Ueki}}, \bibinfo {author} {\bibfnamefont
  {H.}~\bibnamefont {Kimura}}, \bibinfo {author} {\bibfnamefont
  {Y.}~\bibnamefont {Endoh}}, \bibinfo {author} {\bibfnamefont
  {S.}~\bibnamefont {Hosoya}}, \bibinfo {author} {\bibfnamefont
  {G.}~\bibnamefont {Shirane}}, \bibinfo {author} {\bibfnamefont {R.~J.}\
  \bibnamefont {Birgeneau}}, \bibinfo {author} {\bibfnamefont {M.}~\bibnamefont
  {Greven}}, \bibinfo {author} {\bibfnamefont {M.~A.}\ \bibnamefont {Kastner}},
  \ and\ \bibinfo {author} {\bibfnamefont {Y.~J.}\ \bibnamefont {Kim}},\
  }\href@noop {} {\bibfield  {journal} {\bibinfo  {journal} {Phys. Rev. B}\
  }\textbf {\bibinfo {volume} {57}},\ \bibinfo {pages} {6165} (\bibinfo {year}
  {1998})}\BibitemShut {NoStop}%
\bibitem [{\citenamefont {Kofu}\ \emph {et~al.}(2009)\citenamefont {Kofu},
  \citenamefont {Lee}, \citenamefont {Fujita}, \citenamefont {Kang},
  \citenamefont {Eisaki},\ and\ \citenamefont {Yamada}}]{kofu09}%
  \BibitemOpen
  \bibfield  {author} {\bibinfo {author} {\bibfnamefont {M.}~\bibnamefont
  {Kofu}}, \bibinfo {author} {\bibfnamefont {S.-H.}\ \bibnamefont {Lee}},
  \bibinfo {author} {\bibfnamefont {M.}~\bibnamefont {Fujita}}, \bibinfo
  {author} {\bibfnamefont {H.-J.}\ \bibnamefont {Kang}}, \bibinfo {author}
  {\bibfnamefont {H.}~\bibnamefont {Eisaki}}, \ and\ \bibinfo {author}
  {\bibfnamefont {K.}~\bibnamefont {Yamada}},\ }\href@noop {} {\bibfield
  {journal} {\bibinfo  {journal} {Phys. Rev. Lett.}\ }\textbf {\bibinfo
  {volume} {102}},\ \bibinfo {pages} {047001} (\bibinfo {year}
  {2009})}\BibitemShut {NoStop}%
\bibitem [{\citenamefont {Damascelli}\ \emph {et~al.}(2003)\citenamefont
  {Damascelli}, \citenamefont {Hussain},\ and\ \citenamefont
  {Shen}}]{damascelli04}%
  \BibitemOpen
  \bibfield  {author} {\bibinfo {author} {\bibfnamefont {A.}~\bibnamefont
  {Damascelli}}, \bibinfo {author} {\bibfnamefont {Z.}~\bibnamefont {Hussain}},
  \ and\ \bibinfo {author} {\bibfnamefont {Z.}~\bibnamefont {Shen}},\
  }\href@noop {} {\bibfield  {journal} {\bibinfo  {journal} {Rev. Mod.Phys.}\
  }\textbf {\bibinfo {volume} {75}},\ \bibinfo {pages} {473} (\bibinfo {year}
  {2003})}\BibitemShut {NoStop}%
\bibitem [{\citenamefont {Gor'kov}(2013)}]{gorkov13}%
  \BibitemOpen
  \bibfield  {author} {\bibinfo {author} {\bibfnamefont {L.~P.}\ \bibnamefont
  {Gor'kov}},\ }\href@noop {} {\bibfield  {journal} {\bibinfo  {journal} {Phys.
  Rev. B}\ }\textbf {\bibinfo {volume} {88}},\ \bibinfo {pages} {041104}
  (\bibinfo {year} {2013})}\BibitemShut {NoStop}%
\bibitem [{\citenamefont {Yoshida}\ \emph {et~al.}(2003)\citenamefont
  {Yoshida}, \citenamefont {Zhou}, \citenamefont {Sasagawa}, \citenamefont
  {Yang}, \citenamefont {Bogdanov}, \citenamefont {Lanzara}, \citenamefont
  {Hussain}, \citenamefont {Mizokawa}, \citenamefont {Fujimori}, \citenamefont
  {Eisaki}, \citenamefont {Shen}, \citenamefont {Kakeshita},\ and\
  \citenamefont {Uchida}}]{yoshida03}%
  \BibitemOpen
  \bibfield  {author} {\bibinfo {author} {\bibfnamefont {T.}~\bibnamefont
  {Yoshida}}, \bibinfo {author} {\bibfnamefont {X.~J.}\ \bibnamefont {Zhou}},
  \bibinfo {author} {\bibfnamefont {T.}~\bibnamefont {Sasagawa}}, \bibinfo
  {author} {\bibfnamefont {W.~L.}\ \bibnamefont {Yang}}, \bibinfo {author}
  {\bibfnamefont {P.~V.}\ \bibnamefont {Bogdanov}}, \bibinfo {author}
  {\bibfnamefont {A.}~\bibnamefont {Lanzara}}, \bibinfo {author} {\bibfnamefont
  {Z.}~\bibnamefont {Hussain}}, \bibinfo {author} {\bibfnamefont
  {T.}~\bibnamefont {Mizokawa}}, \bibinfo {author} {\bibfnamefont
  {A.}~\bibnamefont {Fujimori}}, \bibinfo {author} {\bibfnamefont
  {H.}~\bibnamefont {Eisaki}}, \bibinfo {author} {\bibfnamefont {Z.~X.}\
  \bibnamefont {Shen}}, \bibinfo {author} {\bibfnamefont {T.}~\bibnamefont
  {Kakeshita}}, \ and\ \bibinfo {author} {\bibfnamefont {S.}~\bibnamefont
  {Uchida}},\ }\href@noop {} {\bibfield  {journal} {\bibinfo  {journal} {Phys.
  Rev. Lett.}\ }\textbf {\bibinfo {volume} {91}},\ \bibinfo {pages} {027001}
  (\bibinfo {year} {2003})}\BibitemShut {NoStop}%
\bibitem [{\citenamefont {Ghiringhelli}\ \emph {et~al.}(2012)\citenamefont
  {Ghiringhelli}, \citenamefont {Tacon}, \citenamefont {Minola}, \citenamefont
  {Blanco-Canosa}, \citenamefont {Mazzoli}, \citenamefont {Brookes},
  \citenamefont {Luca}, \citenamefont {Frano}, \citenamefont {Hawthorn},
  \citenamefont {He}, \citenamefont {Loew}, \citenamefont {Sala}, \citenamefont
  {Peets}, \citenamefont {Salluzzo}, \citenamefont {Schierle}, \citenamefont
  {Sutarto}, \citenamefont {Sawatzky}, \citenamefont {Weschke}, \citenamefont
  {Keimer},\ and\ \citenamefont {Braicovich}}]{ghiringhelli12}%
  \BibitemOpen
  \bibfield  {author} {\bibinfo {author} {\bibfnamefont {G.}~\bibnamefont
  {Ghiringhelli}}, \bibinfo {author} {\bibfnamefont {M.~L.}\ \bibnamefont
  {Tacon}}, \bibinfo {author} {\bibfnamefont {M.}~\bibnamefont {Minola}},
  \bibinfo {author} {\bibfnamefont {S.}~\bibnamefont {Blanco-Canosa}}, \bibinfo
  {author} {\bibfnamefont {C.}~\bibnamefont {Mazzoli}}, \bibinfo {author}
  {\bibfnamefont {N.~B.}\ \bibnamefont {Brookes}}, \bibinfo {author}
  {\bibfnamefont {G.~M.~D.}\ \bibnamefont {Luca}}, \bibinfo {author}
  {\bibfnamefont {A.}~\bibnamefont {Frano}}, \bibinfo {author} {\bibfnamefont
  {D.~G.}\ \bibnamefont {Hawthorn}}, \bibinfo {author} {\bibfnamefont
  {F.}~\bibnamefont {He}}, \bibinfo {author} {\bibfnamefont {T.}~\bibnamefont
  {Loew}}, \bibinfo {author} {\bibfnamefont {M.~M.}\ \bibnamefont {Sala}},
  \bibinfo {author} {\bibfnamefont {D.~C.}\ \bibnamefont {Peets}}, \bibinfo
  {author} {\bibfnamefont {M.}~\bibnamefont {Salluzzo}}, \bibinfo {author}
  {\bibfnamefont {E.}~\bibnamefont {Schierle}}, \bibinfo {author}
  {\bibfnamefont {R.}~\bibnamefont {Sutarto}}, \bibinfo {author} {\bibfnamefont
  {G.~A.}\ \bibnamefont {Sawatzky}}, \bibinfo {author} {\bibfnamefont
  {E.}~\bibnamefont {Weschke}}, \bibinfo {author} {\bibfnamefont
  {B.}~\bibnamefont {Keimer}}, \ and\ \bibinfo {author} {\bibfnamefont
  {L.}~\bibnamefont {Braicovich}},\ }\href@noop {} {\bibfield  {journal}
  {\bibinfo  {journal} {Science}\ }\textbf {\bibinfo {volume} {337}},\ \bibinfo
  {pages} {821} (\bibinfo {year} {2012})}\BibitemShut {NoStop}%
\bibitem [{\citenamefont {Chang}\ \emph {et~al.}(2011)\citenamefont {Chang},
  \citenamefont {Blackburn}, \citenamefont {Holmes}, \citenamefont
  {Christensen}, \citenamefont {Larsen}, \citenamefont {Mesot}, \citenamefont
  {Liang}, \citenamefont {Bonn}, \citenamefont {Hardy}, \citenamefont
  {Watenphul}, \citenamefont {Zimmermann}, \citenamefont {Forgan},\ and\
  \citenamefont {Hayden}}]{chang12}%
  \BibitemOpen
  \bibfield  {author} {\bibinfo {author} {\bibfnamefont {J.}~\bibnamefont
  {Chang}}, \bibinfo {author} {\bibfnamefont {E.}~\bibnamefont {Blackburn}},
  \bibinfo {author} {\bibfnamefont {A.~T.}\ \bibnamefont {Holmes}}, \bibinfo
  {author} {\bibfnamefont {N.~B.}\ \bibnamefont {Christensen}}, \bibinfo
  {author} {\bibfnamefont {J.}~\bibnamefont {Larsen}}, \bibinfo {author}
  {\bibfnamefont {J.}~\bibnamefont {Mesot}}, \bibinfo {author} {\bibfnamefont
  {R.}~\bibnamefont {Liang}}, \bibinfo {author} {\bibfnamefont {D.~A.}\
  \bibnamefont {Bonn}}, \bibinfo {author} {\bibfnamefont {W.~N.}\ \bibnamefont
  {Hardy}}, \bibinfo {author} {\bibfnamefont {A.}~\bibnamefont {Watenphul}},
  \bibinfo {author} {\bibfnamefont {M.}~\bibnamefont {Zimmermann}}, \bibinfo
  {author} {\bibfnamefont {E.~M.}\ \bibnamefont {Forgan}}, \ and\ \bibinfo
  {author} {\bibfnamefont {S.~M.}\ \bibnamefont {Hayden}},\ }\href@noop {}
  {\bibfield  {journal} {\bibinfo  {journal} {Nat. Phys.}\ }\textbf {\bibinfo
  {volume} {8}},\ \bibinfo {pages} {871} (\bibinfo {year} {2011})}\BibitemShut
  {NoStop}%
\bibitem [{\citenamefont {Tabis}\ \emph {et~al.}()\citenamefont {Tabis},
  \citenamefont {Li}, \citenamefont {Tacon}, \citenamefont {Braicovich},
  \citenamefont {Kreyssig}, \citenamefont {Minola}, \citenamefont {Dellea},
  \citenamefont {Weschke}, \citenamefont {Veit}, \citenamefont {Goldman},
  \citenamefont {Schmitt}, \citenamefont {Ghiringhelli}, \citenamefont
  {Bari\v{s}i\'c}, \citenamefont {Chan}, \citenamefont {Dorow}, \citenamefont
  {Yu}, \citenamefont {Zhao}, \citenamefont {Keimer},\ and\ \citenamefont
  {Greven}}]{tabis14}%
  \BibitemOpen
  \bibfield  {author} {\bibinfo {author} {\bibfnamefont {W.}~\bibnamefont
  {Tabis}}, \bibinfo {author} {\bibfnamefont {Y.}~\bibnamefont {Li}}, \bibinfo
  {author} {\bibfnamefont {M.~L.}\ \bibnamefont {Tacon}}, \bibinfo {author}
  {\bibfnamefont {L.}~\bibnamefont {Braicovich}}, \bibinfo {author}
  {\bibfnamefont {A.}~\bibnamefont {Kreyssig}}, \bibinfo {author}
  {\bibfnamefont {M.}~\bibnamefont {Minola}}, \bibinfo {author} {\bibfnamefont
  {G.}~\bibnamefont {Dellea}}, \bibinfo {author} {\bibfnamefont
  {E.}~\bibnamefont {Weschke}}, \bibinfo {author} {\bibfnamefont
  {M.}~\bibnamefont {Veit}}, \bibinfo {author} {\bibfnamefont {A.~I.}\
  \bibnamefont {Goldman}}, \bibinfo {author} {\bibfnamefont {T.}~\bibnamefont
  {Schmitt}}, \bibinfo {author} {\bibfnamefont {G.}~\bibnamefont
  {Ghiringhelli}}, \bibinfo {author} {\bibfnamefont {N.}~\bibnamefont
  {Bari\v{s}i\'c}}, \bibinfo {author} {\bibfnamefont {M.~K.}\ \bibnamefont
  {Chan}}, \bibinfo {author} {\bibfnamefont {C.}~\bibnamefont {Dorow}},
  \bibinfo {author} {\bibfnamefont {G.}~\bibnamefont {Yu}}, \bibinfo {author}
  {\bibfnamefont {X.}~\bibnamefont {Zhao}}, \bibinfo {author} {\bibfnamefont
  {B.}~\bibnamefont {Keimer}}, \ and\ \bibinfo {author} {\bibfnamefont
  {M.}~\bibnamefont {Greven}},\ }\href@noop {} {\bibinfo  {journal} {In
  Preperation}\ }\BibitemShut {NoStop}%
\bibitem [{\citenamefont {Zhou}\ \emph {et~al.}(2003)\citenamefont {Zhou},
  \citenamefont {Yoshida}, \citenamefont {Lanzara}, \citenamefont {Bogdanov},
  \citenamefont {Kellar}, \citenamefont {Shen}, \citenamefont {Yang},
  \citenamefont {Ronning}, \citenamefont {Sasagawa}, \citenamefont {Kakeshita},
  \citenamefont {Noda}, \citenamefont {Eisaki}, \citenamefont {Uchida},
  \citenamefont {Lin}, \citenamefont {Zhou}, \citenamefont {Xiong},
  \citenamefont {Ti}, \citenamefont {Zhao}, \citenamefont {Fujimori},
  \citenamefont {Hussain},\ and\ \citenamefont {Shen}}]{zhao03}%
  \BibitemOpen
\bibfield  {journal} {  }\bibfield  {author} {\bibinfo {author} {\bibfnamefont
  {X.~J.}\ \bibnamefont {Zhou}}, \bibinfo {author} {\bibfnamefont
  {T.}~\bibnamefont {Yoshida}}, \bibinfo {author} {\bibfnamefont
  {A.}~\bibnamefont {Lanzara}}, \bibinfo {author} {\bibfnamefont {P.~V.}\
  \bibnamefont {Bogdanov}}, \bibinfo {author} {\bibfnamefont {S.~A.}\
  \bibnamefont {Kellar}}, \bibinfo {author} {\bibfnamefont {K.~M.}\
  \bibnamefont {Shen}}, \bibinfo {author} {\bibfnamefont {W.~L.}\ \bibnamefont
  {Yang}}, \bibinfo {author} {\bibfnamefont {F.}~\bibnamefont {Ronning}},
  \bibinfo {author} {\bibfnamefont {T.}~\bibnamefont {Sasagawa}}, \bibinfo
  {author} {\bibfnamefont {T.}~\bibnamefont {Kakeshita}}, \bibinfo {author}
  {\bibfnamefont {T.}~\bibnamefont {Noda}}, \bibinfo {author} {\bibfnamefont
  {H.}~\bibnamefont {Eisaki}}, \bibinfo {author} {\bibfnamefont
  {S.}~\bibnamefont {Uchida}}, \bibinfo {author} {\bibfnamefont {C.~T.}\
  \bibnamefont {Lin}}, \bibinfo {author} {\bibfnamefont {F.}~\bibnamefont
  {Zhou}}, \bibinfo {author} {\bibfnamefont {J.~W.}\ \bibnamefont {Xiong}},
  \bibinfo {author} {\bibfnamefont {W.~X.}\ \bibnamefont {Ti}}, \bibinfo
  {author} {\bibfnamefont {Z.~X.}\ \bibnamefont {Zhao}}, \bibinfo {author}
  {\bibfnamefont {A.}~\bibnamefont {Fujimori}}, \bibinfo {author}
  {\bibfnamefont {Z.}~\bibnamefont {Hussain}}, \ and\ \bibinfo {author}
  {\bibfnamefont {Z.-X.}\ \bibnamefont {Shen}},\ }\href@noop {} {\bibfield
  {journal} {\bibinfo  {journal} {Nature}\ }\textbf {\bibinfo {volume} {423}},\
  \bibinfo {pages} {398} (\bibinfo {year} {2003})}\BibitemShut {NoStop}%
\bibitem [{\citenamefont {Comin}\ \emph {et~al.}(2014)\citenamefont {Comin},
  \citenamefont {Frano}, \citenamefont {Yee}, \citenamefont {Yoshida},
  \citenamefont {Eisaki}, \citenamefont {Schierle}, \citenamefont {Weschke},
  \citenamefont {Sutarto}, \citenamefont {He}, \citenamefont {Soumyanarayanan},
  \citenamefont {He}, \citenamefont {Tacon}, \citenamefont {Elfimov},
  \citenamefont {Hoffman}, \citenamefont {Sawatzky}, \citenamefont {Keimer},\
  and\ \citenamefont {Damascelli}}]{comin14}%
  \BibitemOpen
  \bibfield  {author} {\bibinfo {author} {\bibfnamefont {R.}~\bibnamefont
  {Comin}}, \bibinfo {author} {\bibfnamefont {A.}~\bibnamefont {Frano}},
  \bibinfo {author} {\bibfnamefont {M.~M.}\ \bibnamefont {Yee}}, \bibinfo
  {author} {\bibfnamefont {Y.}~\bibnamefont {Yoshida}}, \bibinfo {author}
  {\bibfnamefont {H.}~\bibnamefont {Eisaki}}, \bibinfo {author} {\bibfnamefont
  {E.}~\bibnamefont {Schierle}}, \bibinfo {author} {\bibfnamefont
  {E.}~\bibnamefont {Weschke}}, \bibinfo {author} {\bibfnamefont
  {R.}~\bibnamefont {Sutarto}}, \bibinfo {author} {\bibfnamefont
  {F.}~\bibnamefont {He}}, \bibinfo {author} {\bibfnamefont {A.}~\bibnamefont
  {Soumyanarayanan}}, \bibinfo {author} {\bibfnamefont {Y.}~\bibnamefont {He}},
  \bibinfo {author} {\bibfnamefont {M.~L.}\ \bibnamefont {Tacon}}, \bibinfo
  {author} {\bibfnamefont {I.~S.}\ \bibnamefont {Elfimov}}, \bibinfo {author}
  {\bibfnamefont {J.~E.}\ \bibnamefont {Hoffman}}, \bibinfo {author}
  {\bibfnamefont {G.~A.}\ \bibnamefont {Sawatzky}}, \bibinfo {author}
  {\bibfnamefont {B.}~\bibnamefont {Keimer}}, \ and\ \bibinfo {author}
  {\bibfnamefont {A.}~\bibnamefont {Damascelli}},\ }\href@noop {} {\bibfield
  {journal} {\bibinfo  {journal} {Science}\ }\textbf {\bibinfo {volume}
  {343}},\ \bibinfo {pages} {390} (\bibinfo {year} {2014})}\BibitemShut
  {NoStop}%
\bibitem [{\citenamefont {da~Silva~Neto}\ \emph {et~al.}(2014)\citenamefont
  {da~Silva~Neto}, \citenamefont {Aynajian}, \citenamefont {Frano},
  \citenamefont {Comin}, \citenamefont {Schierle}, \citenamefont {Weschke},
  \citenamefont {Gyenis}, \citenamefont {Wen}, \citenamefont {Schneeloch},
  \citenamefont {Xu}, \citenamefont {Ono}, \citenamefont {Gu}, \citenamefont
  {Le~Tacon},\ and\ \citenamefont {Yazdani}}]{dasilva14}%
  \BibitemOpen
  \bibfield  {author} {\bibinfo {author} {\bibfnamefont {E.~H.}\ \bibnamefont
  {da~Silva~Neto}}, \bibinfo {author} {\bibfnamefont {P.}~\bibnamefont
  {Aynajian}}, \bibinfo {author} {\bibfnamefont {A.}~\bibnamefont {Frano}},
  \bibinfo {author} {\bibfnamefont {R.}~\bibnamefont {Comin}}, \bibinfo
  {author} {\bibfnamefont {E.}~\bibnamefont {Schierle}}, \bibinfo {author}
  {\bibfnamefont {E.}~\bibnamefont {Weschke}}, \bibinfo {author} {\bibfnamefont
  {A.}~\bibnamefont {Gyenis}}, \bibinfo {author} {\bibfnamefont
  {J.}~\bibnamefont {Wen}}, \bibinfo {author} {\bibfnamefont {J.}~\bibnamefont
  {Schneeloch}}, \bibinfo {author} {\bibfnamefont {Z.}~\bibnamefont {Xu}},
  \bibinfo {author} {\bibfnamefont {S.}~\bibnamefont {Ono}}, \bibinfo {author}
  {\bibfnamefont {G.}~\bibnamefont {Gu}}, \bibinfo {author} {\bibfnamefont
  {M.}~\bibnamefont {Le~Tacon}}, \ and\ \bibinfo {author} {\bibfnamefont
  {A.}~\bibnamefont {Yazdani}},\ }\href@noop {} {\bibfield  {journal} {\bibinfo
   {journal} {Science}\ }\textbf {\bibinfo {volume} {343}},\ \bibinfo {pages}
  {393} (\bibinfo {year} {2014})}\BibitemShut {NoStop}%
\bibitem [{\citenamefont {Bari\v{s}i\'c}\ and\ \citenamefont
  {Bari\v{s}i\'c}(2009)}]{barisic09}%
  \BibitemOpen
  \bibfield  {author} {\bibinfo {author} {\bibfnamefont {S.}~\bibnamefont
  {Bari\v{s}i\'c}}\ and\ \bibinfo {author} {\bibfnamefont {O.~S.}\ \bibnamefont
  {Bari\v{s}i\'c}},\ }\href@noop {} {\bibfield  {journal} {\bibinfo  {journal}
  {Physica B}\ }\textbf {\bibinfo {volume} {404}},\ \bibinfo {pages} {370}
  (\bibinfo {year} {2009})}\BibitemShut {NoStop}%
\bibitem [{\citenamefont {Kondo}\ \emph {et~al.}(2013)\citenamefont {Kondo},
  \citenamefont {Palczewski}, \citenamefont {Hamaya}, \citenamefont {Takeuchi},
  \citenamefont {Wen}, \citenamefont {Xu}, \citenamefont {Gu},\ and\
  \citenamefont {Kaminski}}]{kondo13}%
  \BibitemOpen
  \bibfield  {author} {\bibinfo {author} {\bibfnamefont {T.}~\bibnamefont
  {Kondo}}, \bibinfo {author} {\bibfnamefont {A.~D.}\ \bibnamefont
  {Palczewski}}, \bibinfo {author} {\bibfnamefont {Y.}~\bibnamefont {Hamaya}},
  \bibinfo {author} {\bibfnamefont {T.}~\bibnamefont {Takeuchi}}, \bibinfo
  {author} {\bibfnamefont {J.~S.}\ \bibnamefont {Wen}}, \bibinfo {author}
  {\bibfnamefont {Z.~J.}\ \bibnamefont {Xu}}, \bibinfo {author} {\bibfnamefont
  {G.}~\bibnamefont {Gu}}, \ and\ \bibinfo {author} {\bibfnamefont
  {A.}~\bibnamefont {Kaminski}},\ }\href@noop {} {\bibfield  {journal}
  {\bibinfo  {journal} {Phys. Rev. Lett.}\ }\textbf {\bibinfo {volume} {111}},\
  \bibinfo {pages} {157003} (\bibinfo {year} {2013})}\BibitemShut {NoStop}%
\bibitem [{\citenamefont {Vishik}\ \emph {et~al.}()\citenamefont {Vishik},
  \citenamefont {Bari\v{s}i\'c}, \citenamefont {Chan}, \citenamefont {Li},
  \citenamefont {Xia}, \citenamefont {Yu}, \citenamefont {Zhao}, \citenamefont
  {Lee}, \citenamefont {Meevasana}, \citenamefont {Devereaux}, \citenamefont
  {Greven},\ and\ \citenamefont {Shen}}]{vishik14}%
  \BibitemOpen
  \bibfield  {author} {\bibinfo {author} {\bibfnamefont {I.~M.}\ \bibnamefont
  {Vishik}}, \bibinfo {author} {\bibfnamefont {N.}~\bibnamefont
  {Bari\v{s}i\'c}}, \bibinfo {author} {\bibfnamefont {M.}~\bibnamefont {Chan}},
  \bibinfo {author} {\bibfnamefont {Y.}~\bibnamefont {Li}}, \bibinfo {author}
  {\bibfnamefont {D.~D.}\ \bibnamefont {Xia}}, \bibinfo {author} {\bibfnamefont
  {G.}~\bibnamefont {Yu}}, \bibinfo {author} {\bibfnamefont {X.}~\bibnamefont
  {Zhao}}, \bibinfo {author} {\bibfnamefont {W.~S.}\ \bibnamefont {Lee}},
  \bibinfo {author} {\bibfnamefont {W.}~\bibnamefont {Meevasana}}, \bibinfo
  {author} {\bibfnamefont {T.~P.}\ \bibnamefont {Devereaux}}, \bibinfo {author}
  {\bibfnamefont {M.}~\bibnamefont {Greven}}, \ and\ \bibinfo {author}
  {\bibfnamefont {Z.~X.}\ \bibnamefont {Shen}},\ }\href@noop {} {\bibinfo
  {journal} {arxiv:1402.2215}\ }\BibitemShut {NoStop}%
\end{thebibliography}%

\end{document}